\documentclass[journal,10pt]{IEEEtran}

\usepackage{url}
\usepackage{breakurl}
\usepackage[breaklinks]{hyperref}

\usepackage[pdftex]{graphicx}
\usepackage[cmex10]{amsmath}
\usepackage{xcolor}
\usepackage{algorithmic}
\usepackage{array}

\newtheorem{prop}{Proposition}

\usepackage{afterpage}
\usepackage[font=footnotesize]{subfig}

\begin{document}

\title{Who is in Control? Practical Physical Layer Attack and Defense for mmWave based Sensing in Autonomous Vehicles
\vspace{5pt}}

\author{Zhi~Sun ~\IEEEmembership{Senior Member,~IEEE},
Sarankumar Balakrishnan ~\IEEEmembership{Student Member,~IEEE},
Lu Su ~\IEEEmembership{Member,~IEEE},
Arupjyoti Bhuyan ~\IEEEmembership{Senior Member,~IEEE},
Pu Wang ~\IEEEmembership{Member,~IEEE},
and Chunming Qiao ~\IEEEmembership{Fellow,~IEEE}
\thanks{Zhi~Sun and Sarankumar Balakrishnan are with the Department of Electrical Engineering, University at Buffalo, Buffalo, NY 14260 USA (e-mail: zhisun@buffalo.edu; sarankum@buffalo.edu).}
\thanks{Lu Su and Chunming Qiao are with the Department of Computer Science and Engineering, University at Buffalo, Buffalo, NY 14260 USA (e-mail: lusu@buffalo.edu; qiao@buffalo.edu).}
\thanks{A. Bhuyan is with the Idaho National Laboratory (INL), Idaho Falls, ID 83402 USA (e-mail: arupjyoti.bhuyan@inl.gov).}
\thanks{P. Wang is with the Department of Computer Science, University of North Carolina at Charlotte, Charlotte, NC 28223 USA (e-mail: pu.wang@uncc.edu).}
\thanks{This work was supported by the INL Laboratory Directed Research and
Development (LDRD) Program under the DOE Idaho Operations Office under
Contract DE-AC07-05ID14517.}
\thanks{This work has been submitted to the IEEE for possible publication. Copyright may be transferred without notice, after which this version may no longer be accessible.}
\vspace{-5pt}
}

\maketitle

\begin{abstract}
With the wide bandwidths in millimeter wave (mmWave) frequency band that results in unprecedented accuracy, mmWave sensing has become vital for many applications, especially in autonomous vehicles (AVs). In addition, mmWave sensing has superior reliability compared to other sensing counterparts such as camera and LiDAR, which is essential for safety-critical driving. Therefore, it is critical to understand the security vulnerabilities and improve the security and reliability of mmWave sensing in AVs. To this end, we perform the end-to-end security analysis of a mmWave-based sensing system in AVs, by designing and implementing practical physical layer attack and defense strategies in a state-of-the-art mmWave testbed and an AV testbed in real-world settings. Various strategies are developed to take control of the victim AV by spoofing its mmWave sensing module, including adding fake obstacles at arbitrary locations and faking the locations of existing obstacles. Five real-world attack scenarios are constructed to spoof the victim AV and force it to make dangerous driving decisions leading to a fatal crash. Field experiments are conducted to study the impact of the various attack scenarios using a Lincoln MKZ-based AV testbed, which validate that the attacker can indeed assume control of the victim AV to compromise its security and safety. To defend the attacks, we design and implement a challenge-response authentication scheme and a RF fingerprinting scheme to reliably detect aforementioned spoofing attacks.
\vspace{-3pt}
\end{abstract}


%

\section{Introduction}
Autonomous vehicles (AVs) are envisioned to be the future of transportation. It is predicted that in the coming decades (by 2045) more than half of the new vehicles manufactured will be autonomous \cite{litman2017autonomous}. 
Fully autonomous vehicles need human-like cognition capabilities to safely navigate the environment and react to unforeseen circumstances. To achieve real-time human-like sensing capability, AVs rely on multitude of heterogeneous sensors such as camera, LiDAR, and radar. 
Millimeter wave (mmWave) radars have become an attractive choice in AVs due to their better spatial resolution thanks to the wide bandwidth available at mmWave spectrum. Currently, the mmWave sensor is a crucial component of state-of-the-art software systems for AVs such as Baidu Apollo \cite{ApolloBaidu} and Autoware \cite{kato2018autoware}. Open-source software systems like \textit{OpenPilot} uses mmWave radar as a major component for its lane keep assist and forward collision warning capabilities\cite{openpilot}. 

\par
With increasing companies vying for the lucrative AV market and the predicted prevalence of AVs on the road, the safety-critical operation of AVs is important. AV safety has been a subject of intense scrutiny over the past few years. Even though AVs use state-of-the-art sensors and software systems to make reliable decisions and perform safe driving actions, multiple serious or fatal accidents involving AVs have been reported \cite{teslaAccidentMay2016}, \cite{Tesla2018Accident}, \cite{Teslaaccident}. 
Those accidents show that sensors and software used in AVs are prone to vulnerabilities, which raises a serious question: \textit{\textbf{Can a powerful adversary take advantage of the vulnerabilities of the sensors used in AVs and cause an intentional safety-critical incident?}} For example, the camera in the AV could be attacked by blinding it with laser beams thus derailing the AVs capabilities to detect lanes and traffic signs leading to severe consequences \cite{petit2015remote}. Similarly, LiDAR could be spoofed with fake obstacles which could be perceived as a vehicle thus spoofing the victim AV to make erroneous critical decisions which may lead to fatal crash \cite{shin2017illusion}, \cite{cao2019adversarial}. 

Despite the important role of mmWave sensing in AVs, the understanding of its vulnerabilities is far from sufficient. 
In \cite{chauhan2014platform}, attacks aiming to spoof chirp-based radar are investigated using software defined radios. However, the attack is highly impractical since the attacker uses a physical cable connecting to the victim's system to send the spoofing signals. The radar also only works at 2.4GHz, not the mmWave band. 
In \cite{yan2016can}, a preliminary attempt to spoof mmWave sensor in AVs using waveform generators is reported. However, the designed spoof signals do not use the chirp waveform that is used in mmWave radar. Instead, the waveform generator only generates simple noise signals that can be easily detected by the victim.
In summary, the attacks developed in existing works are too simple to cause meaningful security consequences in AVs.

\par
To our best knowledge, no existing work performs comprehensive security analysis of the AV mmWave sensor module and its implications at driving decision level in real-world driving scenarios. On that front, this paper aims at performing practical physical layer security analysis of mmWave radar in AVs by designing and implementing practical attack and defense strategies. We ask the following questions: \textit{\textbf{Is it possible to reliably spoof an mmWave sensor used in the victim AV? And in doing so, will the attacker be able to cause meaningful security consequences to the victim AV?}}.

\par 
There are three key challenges in answering the above questions. 
(a) \textit{Is it feasible to spoof a mmWave sensor to perceive a fake obstacle?} The attacker must be able to generate the replica of the signal used by the victim mmWave sensor, with a precisely controlled time delay. 
(b) \textit{Can the attacker continuously spoof an obstacle to deceive the AV?} 
The attacker needs to continuously spoof the AV to influence the AV's decision making process. Hence, the attacker should continuously track the position of the victim AV to update the time delay with which the attacker must spoof. 
More importantly, in addition to (a) and (b), we ask a more ambitious question.
(c) \textit{\textbf{Is it possible for the attacker to make the AV crash on to a lead obstacle by deceiving the victim AV to perceive the lead obstacle out of the danger zone?}} To deceive the AV into recognizing the lead vehicle as out of the danger zone, for instance, perceiving it as an obstacle in adjacent lane, the attacker must deviate the obstacle from the real position to a strategically controlled position, which is a daunting task. 

\par
In this work, we solve all the above challenges and show that the mmWave sensor in the victim AV can be reliably spoofed using a multi-attacker approach, leading to severe security consequences. 
First, to address (a), we develop a mmWave radar adversary system using the state-of-the-art mmWave software defined radio \cite{niSDR}, which is capable of transmitting a replica of the waveform used by the victim AV. 
Second, to address (b), we design a tracking module to continuously track the position and velocity of the victim AV and update the delay parameters of the spoofing signals.
Finally, to address (c), we identify the challenges in faking the locations of existing obstacles and develop three sophisticated strategies: (1) random deviation of an obstacle by using Gaussian signal attack; (2) synchronous attacks that can fake the positions of obstacles in a controlled manner, and (3) since the perfect synchronization among distributed attackers is not practical, we develop a novel asynchronous attack strategy, which explores the signal correlation in the position estimation method used in the mmWave sensor. The attack strategy can reliably fake the locations of existing obstacles from their real positions to the attacker controlled strategic positions. 
We validate the developed strategies by conducting real-world drive-by experiments and perform end-to-end security analysis using Baidu Apollo \footnote{Baidu Apollo is an open-source software that implements perception, planning, and control for AVs and is widely used by AV industry.} based AV software \cite{ApolloBaidu}. 

\par 
To understand the severity of the security impact of our developed attacks, we construct 5 real-world attack scenarios: 1) \textit{AV Stalling Attack,} which creates traffic chaos; 2) \textit{Hard Braking Attack,} which endangers the safety of the passengers and the vehicle behind; 3) \textit{Lane Changing Attack,} which forces the victim AV into making unintentional lane change; 4) \textit{Multi-stage Attack,} which combines multiple attack stages with the goal of leading the AV to a fatal crash; and 
5) \textit{Cruise Control Attack,} which spoofs the cruise control function of the victim AV and leads to a fatal crash.

\par 
As defense mechanism, we propose and implement two physical layer solutions to detect the developed spoofing attacks. First, we propose a \textit{challenge-response} defense where the victim AV detects the spoofing attacks from the response to the transmitted randomized challenge waveform. Second, we adopt an RF fingerprinting scheme to distinguish spoofed received signal from the legitimate received signal, thereby detecting the spoofing attacks. 
\par 
Our major contributions are summarised as follows:
\begin{itemize}
    \item We perform the first comprehensive physical layer security analysis of AV mmWave sensing. Practical spoofing attack strategies are developed and implemented, which are proved to be able to cause fatal consequences in real world. The proposed attacking strategies are not just heuristic but based on rigorous mathematical analysis on the state-of-the-art AV sensing algorithms, which has guaranteed spoofing performance.
    \item We develop a software defined radio based mmWave testbed to perform sophisticated single node and multi-node spoofing attacks on the mmWave sensor module in a Lincoln MKZ-based AV testbed. 
    \item To study the severity of the developed attack on the AC security, we conduct extensive real-world experiments on the Lincoln MKZ AV fitted with state-of-the-art sensors. Five attack scenarios are designed to impart various levels of security impacts on the AV. 
    \item We show that our spoofing attacks impact the end-to-end security of AVs and spoof the AV into making hazardous safety-critical driving decisions.
    \item As a defense against the spoofing attack, we propose and implement two spoofing attack mitigation strategies, including the \textit{challenge-response} authentication and the RF fingerprinting mechanism.
\end{itemize}

\section{Related Work}
\textbf{Sensor attacks.} In \cite{yan2016can}, jamming and spoofing attacks are performed on ultrasonic sensors. Blinding attacks on camera and preliminary jamming attacks on mmWave radars are also investigated. In \cite{petit2015remote}, physical sensor attack are performed on AV camera and LiDAR. In \cite{cao2019adversarialOld} and \cite{cao2019adversarial}, adversarial attack are launched on LiDAR-based perception in AVs. In \cite{ranganathan2012physical}, an attack strategy is developed to spoof chirp-based ranging system. In \cite{chauhan2014platform} and \cite{chauhan2014demonstration}, the feasibility of spoofing FMCW based radar is explored. However, the attacker has to use physical cables connected to the victim radar to launch the attack. In \cite{miura2019low}, a distance spoofing attack on mmWave FMCW radar is presented. Contrary to all those works, we perform the end-to-end security analysis of AV mmWave sensing system in real AV testbed in various practical real-world scenarios.

\par
\textbf{System level attacks.} In \cite{jha2019kayotee} and \cite{jha2019ml}, fault injection tools are proposed to study the safety and reliability of various AV hardware and software components. In \cite{rubaiyat2018experimental}, the resilience of \textit{OpenPilot}, an open-source Adaptive cruise control (ACC), LKAS, and Assisted lane change system for AVs is evaluated, where radar is used as a primary sensor. 
Contrary to the security analysis using synthetic data, in this work, we evaluate the end-to-end security vulnerability of the AV by performing physical spoofing attack on the victim AV.

\par
\textbf{Adversarial attacks.} Recently, significant efforts have been made to investigate the security of AVs against adversarial sensor attacks \cite{cao2019adversarial}, \cite{sitawarin2018darts}, \cite{chernikova2019self}, \cite{jia2019fooling}, \cite{boloor2020attacking}. Those works target the machine learning model used in AVs, whereas this paper focuses on the physical attack of the radar used in AVs.

\begin{figure*}[htbp]
    \centering
    \includegraphics[width=0.9\textwidth]{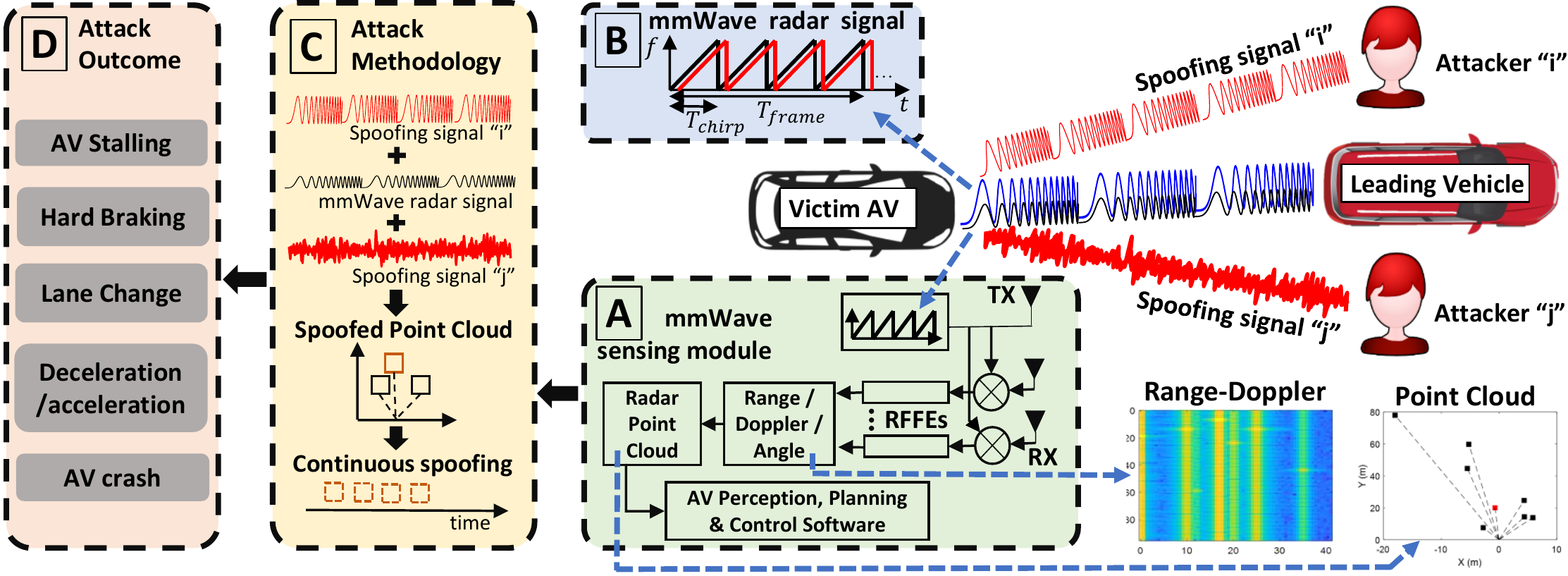}
    \vspace{5pt}
    \caption{Illustration of the adversarial setting and victim AV system overview. Block [A] shows the mmWave sensing module used in AVs. Block [B] shows the transmitted and received mmWave chirp waveform. Block [C] illustrates the attack methodology. Block [D] shows the driving decisions taken by the AV due to the attack described in block [C].}
    \vspace{10pt}
    \label{fig:systemoverview}
\end{figure*}

\par
\textbf{Defense.} The integrity of sensor signals can be verified at the physical layer by considering the physics governing the sensor. For instance, the physical propagation characteristics of sensor signals can be exploited to design attack detection mechanisms. In \cite{shoukry2015pycra}, PyCRA, a physical challenge-response authentication scheme is proposed to defend magnetic and RFID sensors against malicious spoofing attacks. In this work, we propose to use (1) the \textit{challenge-response} mechanism that randomizes the waveform parameters of the mmWave radar, as well as (2) the \textit{RF fingerprinting} mechanism that exploits signal characteristics, to defend against malicious attacks.

\section{System Overview And Adversary Model}

In this section, we overview the proposed attack system, introduce the background of the mmWave sensing module in AV, and discuss the threat model and attack goals.
Fig.\ref{fig:systemoverview} shows the overview of the attack system architecture and the AV mmWave sensing module. The proposed attack system consists of multiple distributed attackers (Attacker "i" and "j" in Fig.\ref{fig:systemoverview}) transmitting spoofing signals with the objective of deceiving the victim AV (black car in Fig.\ref{fig:systemoverview}) into making dangerous decisions. The mmWave sensing module of the victim AV is shown in block [A] of Fig. \ref{fig:systemoverview}, which is introduced in Sec. \ref{sec: radar sensing}. 
The objective of the attackers is to spoof a reliable fake obstacle at an arbitrary location or fake the location of an existing obstacle (e.g., the leading vehicle), using the methods discussed in Sec. \ref{sec:attackdesign}.
The key methodology is to generate a spoofed point cloud (i.e., the positioning results derived by the mmWave radar) by combining the legitimate mmWave radar signal (block [B]) with the spoofing signals, as shown in block [C] in Fig. \ref{fig:systemoverview}.
Such spoofing attacks cause the victim AV to make dangerous safety-critical decisions as shown in block [D] in Fig. \ref{fig:systemoverview}. The detailed attacker goals and assumptions in our threat model are discussed in Sec. \ref{sec: Adversary Model}.

\subsection{Background: mmWave radar module in AV}\label{sec: radar sensing}
A mmWave radar has one transmitter and multiple receivers, as shown in block [A] in Fig. \ref{fig:systemoverview}. The transmitter generates the chirp waveform (block [B] in Fig. \ref{fig:systemoverview}), which is:
\begin{equation}\label{eq:xt}
x(t)=\cos\left(2\pi f_{start} t + \frac{\pi B}{T_{chirp}}t^2\right),
\end{equation}
where $f_{start}$ is the transmission start frequency, $B$ is the sweep bandwidth, i.e., $f_{end}-f_{start}$, and $T_{chirp}$ is the chirp duration. 
Then the chirp signal is reflected back by the obstacle at distance $d$ and received by the multiple receivers at the mmWave radar. 
The received signal is given by
\begin{equation}\label{eq:rt}
 r(t)\!=\!\alpha \cos\!\left(\!2\pi f_{start} (t\!-\!t_{delay}) + \frac{\pi B}{T_{chirp}}(t\!-\!t_{delay})^2\!\right)\!,   
\end{equation}
where $t_{delay} = \frac{2(d+vt)}{\lambda}$ is the time delay due to an obstacle at distance $d$ moving with velocity $v$ with respect to the AV. This received signal is mixed with the transmitted signal in Eq. \ref{eq:xt}, which produces a cosine signal given by 
\begin{equation}\label{eq:fb}
y_{mixed}(t)\!=\!\cos\!\left(\!2\pi f_{start}t_{delay}-\pi\frac{B}{T}t_{delay}^2-2\pi\frac{B}{T}t_{delay} t\!\right)\!,
\end{equation}
where $\frac{B}{T}\tau$ is the beat frequency $f_{b}$, which is proportional to the distance $d$. The beat frequency can be calculated by $f_b = \frac{2RB}{cT_{chirp}}+f_{doppler}$ with $f_{doppler}=\frac{2vcos\psi}{\lambda}$.

\par The mixed signal in Eq. \ref{eq:fb} is further processed by the sensing module shown in block [A] in Fig. \ref{fig:systemoverview}, which consists of three subsystems: \textit{range \& velocity estimation}, \textit{direction estimation}, and \textit{point cloud generation}.

\paragraph*{\textbf{Range \& Velocity Estimation}}
The distance from the mmWave radar to the object is derived by taking the FFT of the signal in Eq. \ref{eq:fb} and finding the $f_{b}$ corresponding to the peak of the FFT. Then the distance is derived by $d = \frac{f_{b}Tc}{2 B}$. The mobility of the obstacle introduces a Doppler shift with the Doppler frequency $f_{doppler} = \frac{2v f_{start}t cos(\theta)}{c}$. Accordingly, in the mobile case, the received signal is given by 
\begin{equation}\label{eq:rxdoppler}
    r(t)\!=\! \cos\!\left(\!2\pi f_{start}(t\!-\!t_{delay}) +\phi_{doppler}+\pi \frac{B}{T}(t\!-\!t_{delay})^2\!\right)\!,
\end{equation}
where $\phi_{doppler}=2\pi f_{doppler}*n*T$ and $n$ is the chirp index in a frame. The velocity of the obstacle is determined by taking second FFT across chirps in different time indices. The range FFT and Doppler FFT result in a Range-Doppler map of dimension $R_{bin} \times V_{bin}$ where $R_{bin}$ is the number of range bins and $V_{bin}$ is the number of velocity bins. An example of the Range-Doppler map is given in Fig. \ref{fig:systemoverview}.
The estimated range bin and velocity bin of each of the obstacle in the Range-Doppler map is denoted as a tuple $(r_{m}^{k},v_{n}^{k})$, where $k$ is the index of the obstacle, $m$ and $n$ are the index of the range bin and velocity bin, respectively.

\paragraph*{\textbf{Direction Estimation}} \label{sec: angle estimation}
As shown in block [A] in Fig. \ref{fig:systemoverview}, an AV mmWave radar uses multiple receiving antennas. 
The Range-Doppler maps of all receiving antennas result in a data cube of dimension $N\times R_{bin} \times V_{bin}$ where $N$ is the number of receiving chains. The direction $\theta$ of each of the obstacle $(r_{m}^{k},v_{n}^{k})$ can be accurately estimated using subspace-based beamforming methods, such as the Multiple Signal Identification and Classification (MUSIC) algorithm 
\cite{schneider2005automotive}. For a mmWave radar with $N$ receiving antennas and arriving signals from $K$ directions, MUSIC algorithm separates $N-K$ noise subspace $Z$ from the $K$ signal subspace $S(\theta_1,\theta_2,\cdots,\theta_K)$ using subspace decomposition techniques such as Eigen decomposition. 
Then, the $K$ arriving signal directions can be estimated by the $K$ minima of function $\frac{1}{a(\theta)'*Z*Z'*a(\theta)}$, where $a(\theta)$ is the steering vector corresponding to the direction $\theta$.

\paragraph*{\textbf{Point Cloud Generation}} \label{sec: Point-Cloud Generation}
The identified positions of the obstacles are presented as a Point Cloud for further processing by the AVs perception, planning, and control software modules. An example of the Point Cloud is given in Fig. \ref{fig:systemoverview}. The details of AV software modules are given in Appendix \ref{sec:softwaremodules}.

\begin{figure*}
    \begin{minipage}[t]{0.23\textwidth}
    \includegraphics[trim={1.5cm 8.5cm 16.5cm 2cm},clip,width=1.65\textwidth]{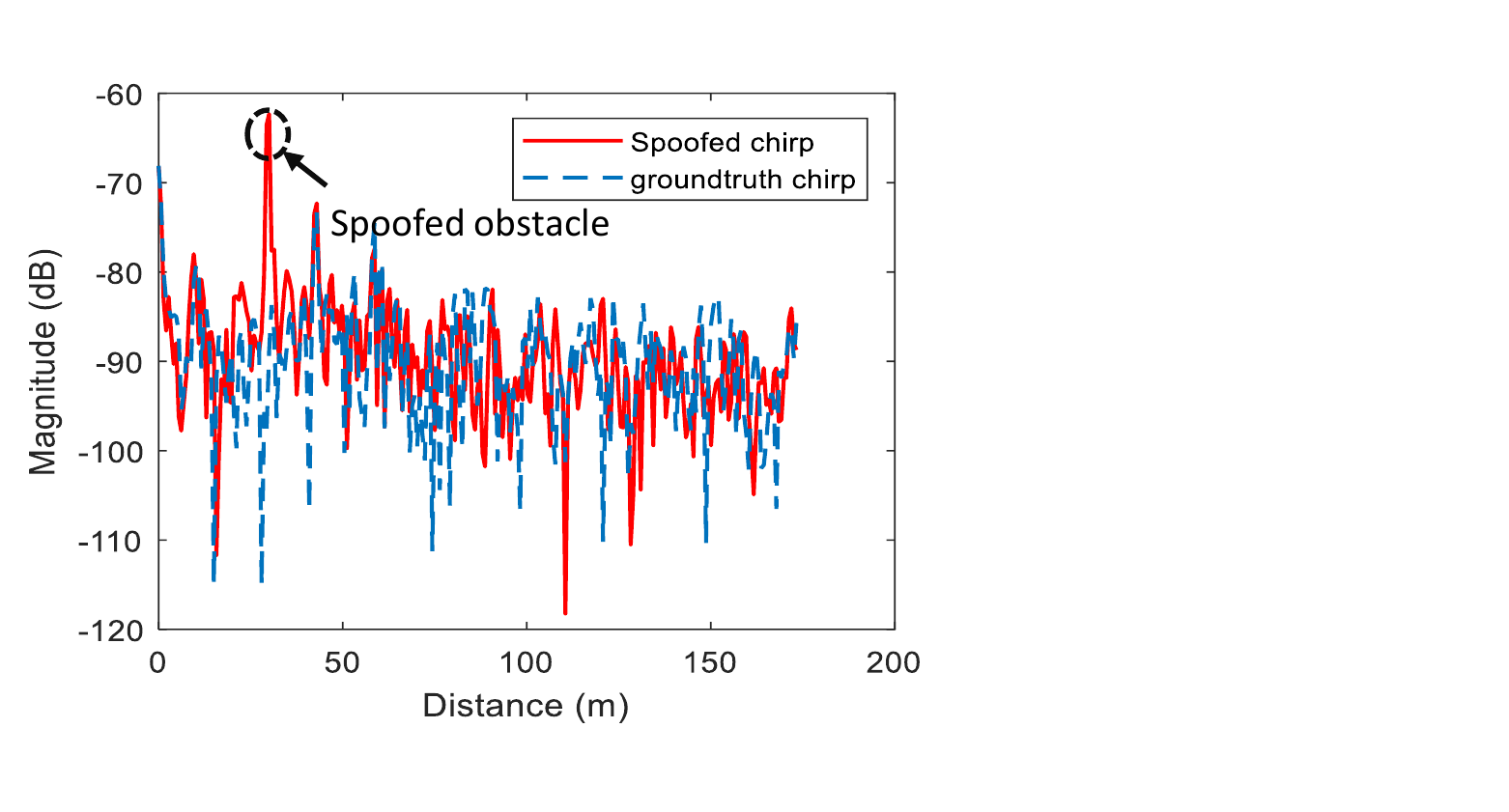}
    \caption{Range FFT of a chirp showing the spoofed obstacle at 30m.}
    \label{fig:distanceSpoof}
    \end{minipage}
    ~~
    \begin{minipage}[t]{0.23\textwidth}
    \includegraphics[trim={1cm 9.5cm 22cm 1.5cm},clip,width=1.15\textwidth]{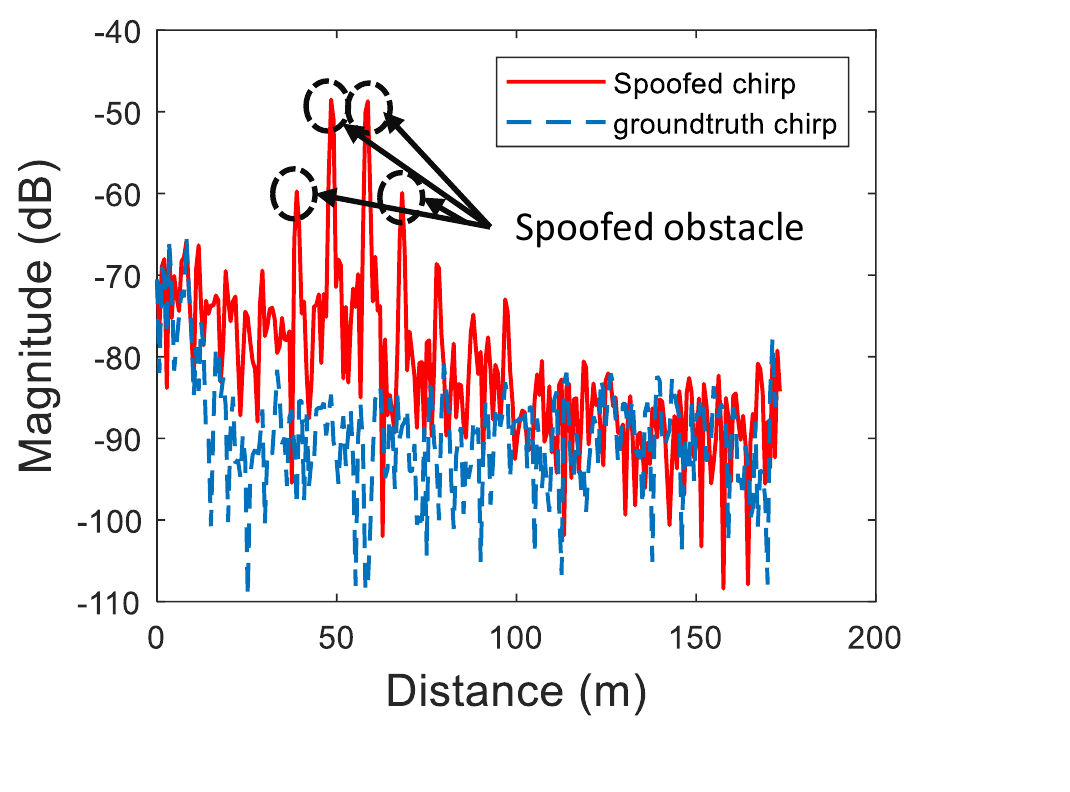}
    \caption{Range FFT of a chirp showing the multiple spoofed obstacle.}
    \label{fig:multipledistanceSpoof}
    \end{minipage}
    ~~
    \begin{minipage}[t]{0.23\textwidth}
    \includegraphics[trim={5.5cm 9cm 6cm 10cm},clip,width=1\textwidth]{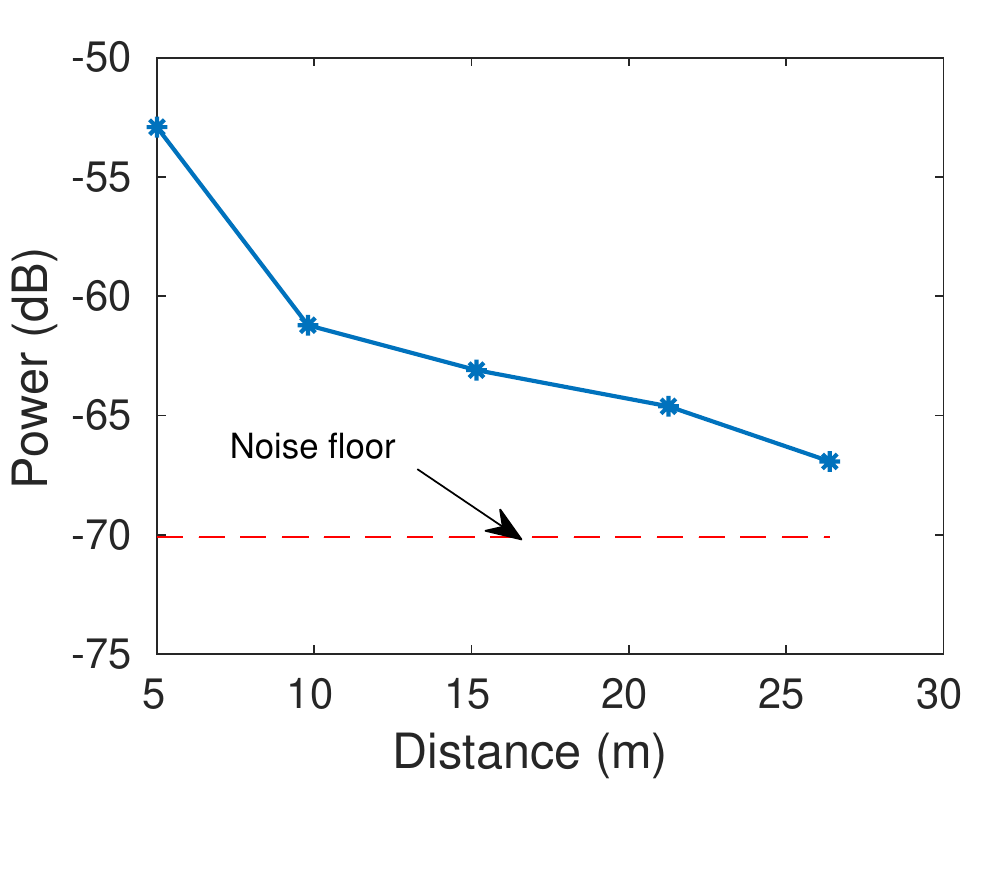}
    \caption{Received power of the spoofing signal as a function of attacker's distance.}
    \label{fig:distance performance}
    \end{minipage}
    ~~
    \begin{minipage}[t]{0.23\textwidth}
    \includegraphics[trim={5.5cm 9cm 6cm 10cm},clip,width=1\textwidth]{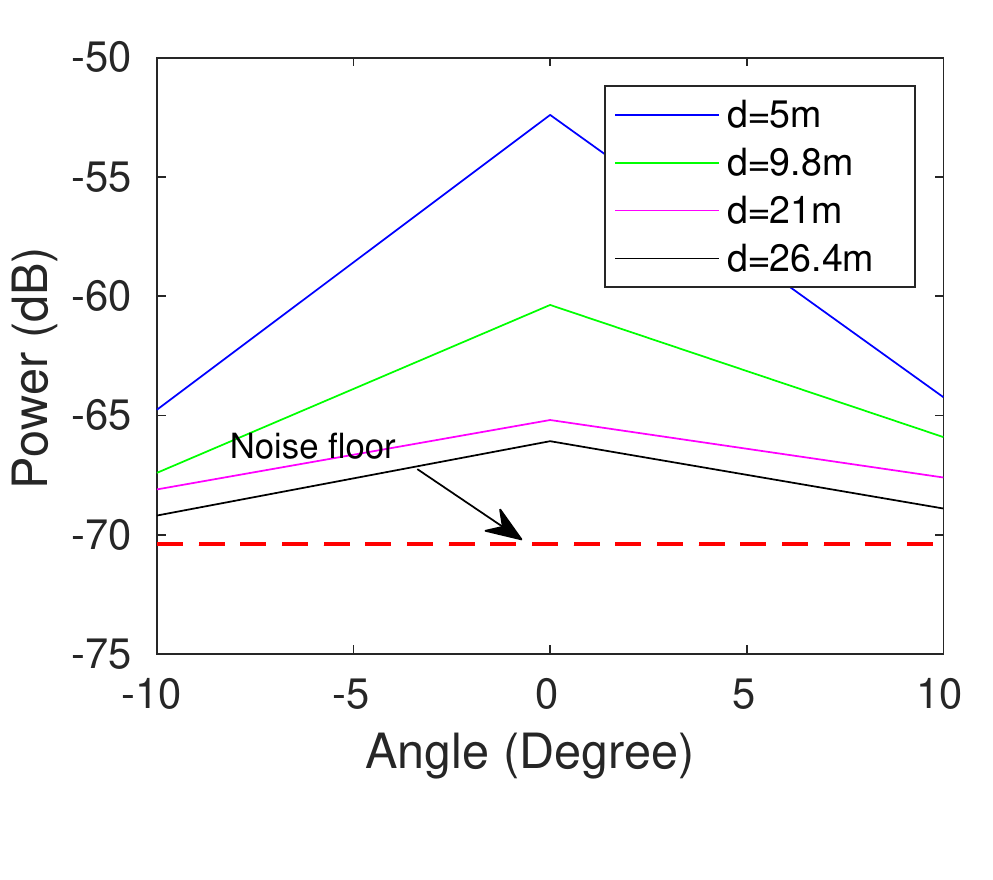}
    \caption{Received power of the spoofing signal as a function of attacker's angle. }
    \label{fig:Angle performance}
    \end{minipage}
    \vspace{6pt}
\end{figure*}

\subsection{Threat Model and Attack Goal}
\label{sec: Adversary Model}

\subsubsection{Threat Model} We identify mmWave radar spoofing as a threat model to spoof the victim AV into making dangerous safety-critical decisions. A multiple-attacker spoofing model are considered, with the following assumptions:

\paragraph*{\textbf{Prior knowledge}} The attacker has prior knowledge on the type of mmWave radar used in victim AV. Hence, the key parameters are available, including the waveform parameters (frequency sweep bandwidth and slope), the number of chirp signals used in a frame, and the duty cycle of the radar frame. The attackers do not need to know the perception algorithm used by the victim AV.

\paragraph*{\textbf{Radar Hardware Physical Access}}
The attackers do not have physical access to the radar hardware, firmware, and software. The attackers can not modify the firmware or inject false data in to the system.

\paragraph*{\textbf{Position of the Attackers}} The attackers are free to take any position. For stealthiness, the attackers in our experiments are stationary and are positioned on the side of the roadways. Also, there are physical signal propagation constraints (discussed in Sec. \ref{sec:attackerangle}) that limits the attackers' positions. 

\paragraph*{\textbf{Attack equipment}} The attackers use a mmWave receiver for sensing the victim radar's signal and a mmWave transmitter for performing replay and spoofing attack. The transmitter and receiver are discussed in Sec. \ref{sec:sdr}. 

\subsubsection{Attack Goal}
The attack goal is to spoof the victim AV into making potentially dangerous driving decisions or actions as shown in block [D] in Fig. \ref{fig:systemoverview}, which endanger the occupants of the AV as well as the other vehicles in the vicinity. 

\vspace{3pt}
\section{Attack Design: Physical Attacks on AV mmWave Radar}
\label{sec:attackdesign}
\vspace{3pt}

In this section, we develop the two key spoofing strategies, including adding fake obstacles at arbitrary locations and faking the locations of existing obstacles, which form the more sophisticated attack scenarios described in Sec. \ref{sec:case study}.

\begin{figure*}
    \begin{minipage}{0.47\textwidth}
    \includegraphics[trim={1cm 9cm 13cm 2cm},clip,width=1.05\textwidth]{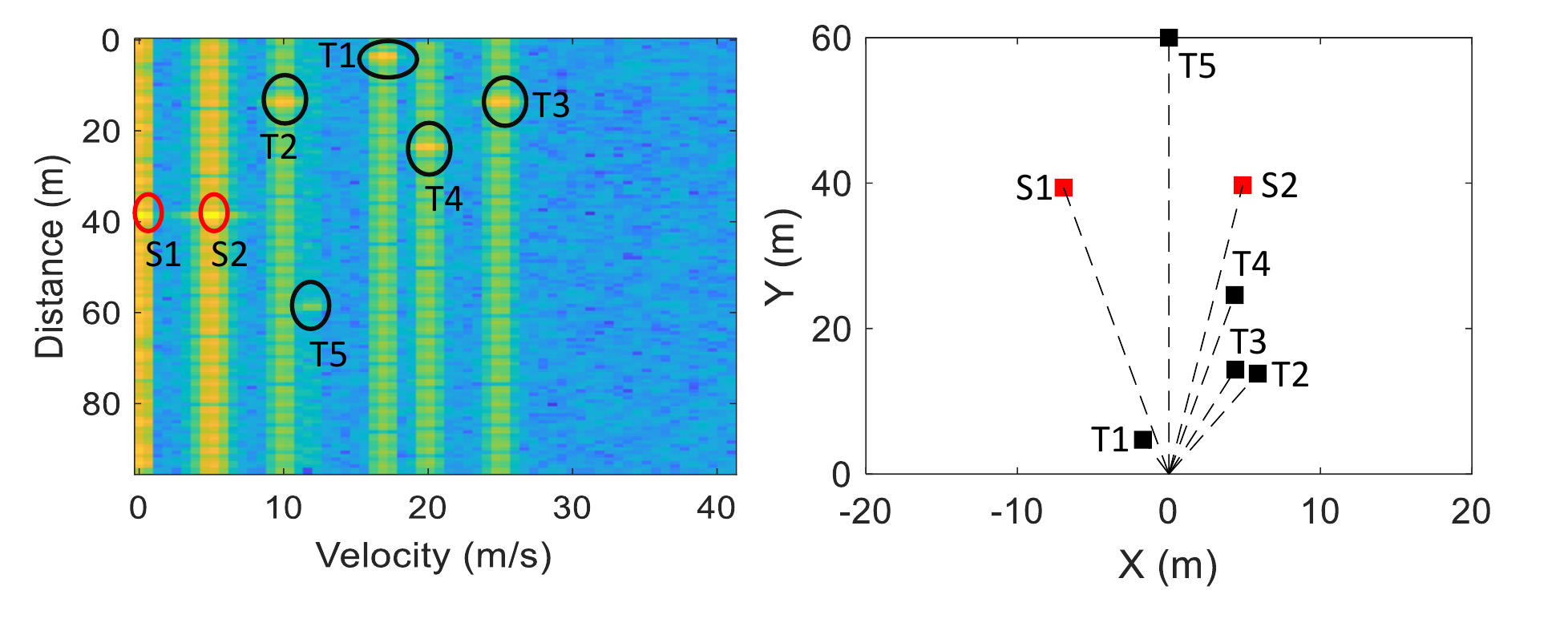}
    \caption{Range-Doppler map and point cloud of the identified obstacles ($T_i$) and the spoofed obstacles ($s1$ and $s2$). The attackers are not synchronized leading to the two spoofing signals $s1$ and $s2$ resolved in Range-Doppler domain. }
    \label{fig:NotSyncAngleAttack}
    \end{minipage}%
    \hspace{1cm}
    \begin{minipage}{0.47\textwidth}
    \includegraphics[trim={1cm 9cm 13cm 2cm},clip,width=1.05\textwidth]{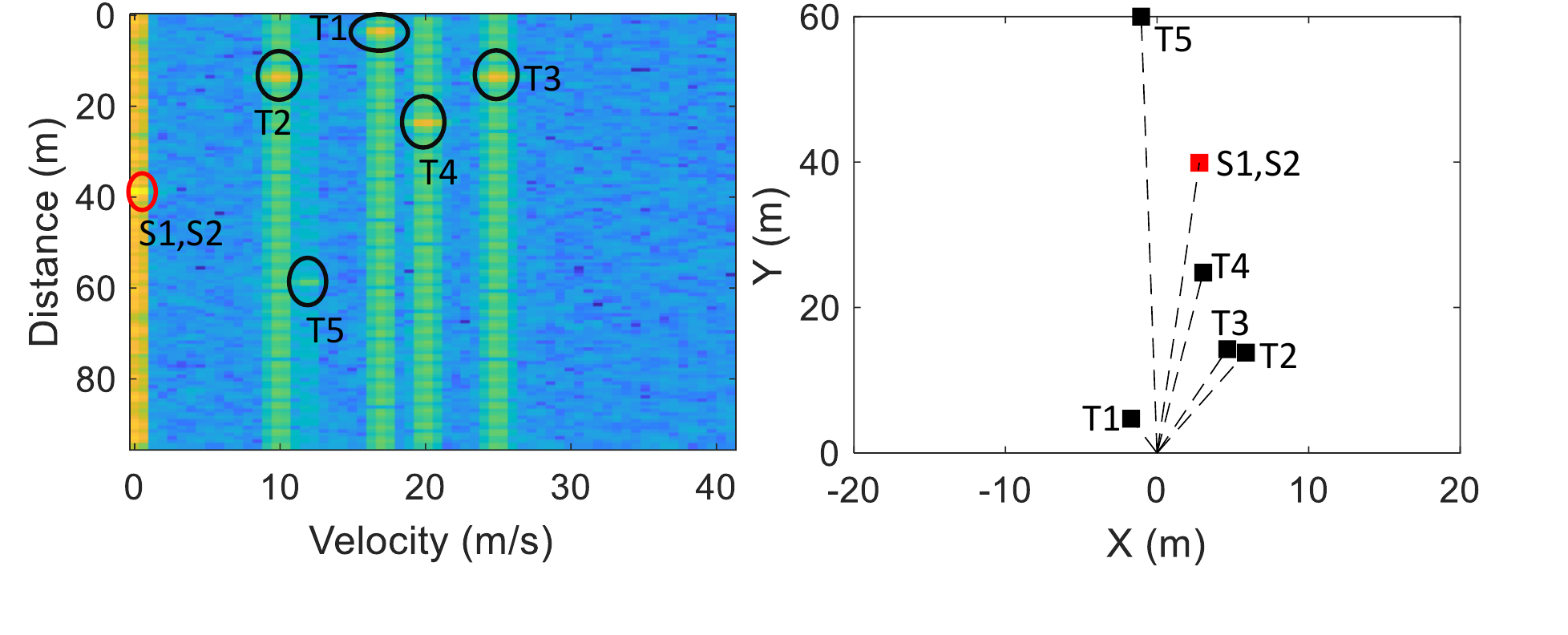}
    \caption{Range-Doppler map and point cloud of the identified obstacles ($T_i$) along with the spoofed obstacle ($s1$ and $s2$). The attackers are perfectly synchronized leading to the two spoofing signals resolved as single obstacle $s1,s2$.}
    \label{fig:SyncAngleAttack}
    \end{minipage}
    \vspace{6pt}
\end{figure*}

\vspace{-6pt}
\subsection{Adding Fake Obstacles}\label{sec:distancespoofing}

The fundamental yet core attack capability of an attacker (for generality, assume "attacker i" in Fig. \ref{fig:systemoverview}) is to reliably trick a victim AV's radar to perceive a fake obstacle spoofed by the attacker as a legitimate obstacle. An AV can be spoofed with a fake obstacle at a particular distance by transmitting a spoofing waveform identical to the waveform in Eq. \ref{eq:xt} used by the victim AV with a carefully designed delay. An attacker could sense the transmitted radar waveform from the AV and either replay a previously recorded waveform or impersonate the transmitted waveform after a delay of $t_{delay} = \frac{2*d}{c}$, where $d$ is the desired distance of the spoofed obstacle from the victim radar. 

\paragraph*{\textbf{Spoofing a mobile obstacle}} \label{sec:attacker distance}
The received waveform reflected by the moving object with velocity $v$ is given by Eq. \ref{eq:rxdoppler}. The velocity progresses linearly with chirp number, which can be estimated by the relationship $f_{doppler} n T_{chirp} = \frac{2v f_{start}}{c} n T_{chirp}$. The attacker can use software defined radios to easily change the phase of its transmitted waveform. If the desired spoofed velocity is $v$, the manipulated phase should be $\frac{2v f_{start}}{c} n T_{chirp}$ for each of the chirp $n$. 
By linearly increasing the chirp phase, the attacker can spoof an object with the fake velocity. As an example, Fig. \ref{fig:distanceSpoof} shows a Range FFT where a spoofed obstacle at 30m is perceived by the victim AV. 

\paragraph*{\textbf{Spoofing multiple obstacles}}
\label{sec:multiplespoof}
The transmitted spoofing signal is a superposition of signals corresponding to various delays $\tau_{delay}$ between the mmWave radar and the multiple spoofed obstacles. The spoofing signal is given by 
\begin{equation}\label{eq:multiplespoof}
x_{spoof}(t)\!=\!\sum_{n=1}^{N}\cos\!\left(2\pi f_c (t\!-\!\tau_{delay,n})\!+\!\frac{\pi B}{T}(t\!-\!\tau_{delay,n})^2\right)\!,    
\end{equation}
where $N$ is the number of spoofed obstacles. For our testbed described in Sec. \ref{sec:sdr}, the delay $\tau_{delay}$ in terms of number of samples $m$ can be calculates as $mT_{s}=\frac{d}{c}$, where $1/T_{s}$ is the sampling rate of the transmitter, and $d$ is the distance between the two spoofed obstacles. As an example, Fig. \ref{fig:multipledistanceSpoof} shows the Range FFT where multiple spoofed obstacles are created. The attacker simultaneously spoofs four obstacles with $10m$ distance between each of them.

\paragraph*{\textbf{Influence of Attacker's Distance}} \label{sec:attacker distance}
Since the attacker need to sense the mmWave signal from the victim AV and transmit the spoofing waveform, the attacker should not be too far away. 
Fig. \ref{fig:distance performance} shows the received power of the attacker's spoofing signal for varying distances between the attacker and the victim radar, when using our mmWave testbed described in Sec. \ref{sec:sdr}. 
The maximum distance from the victim AV that the attacker can still reliably spoof is $26 m$. 
This limitation is due to the limited gain and transmit power of the antenna module of the attacker [see Sec. \ref{sec:attacker system}]. The gain of the antenna used for the attack is $23 dBi$. The attack range can be significantly improved by using a high gain directional antenna \cite{vubiq42dbi}. 

\paragraph*{\textbf{Influence of Attacker's Angle}}
\label{sec:attackerangle}
Since attackers are on road side, there is an angle between the victim AV's moving direction and the attacker antenna's direction. 
The attacker has restricted angular freedom in launching the spoofing attack since the mmWave radars in existing cars have a narrow field-of-view, typically $\pm 10^{\circ}$. Fig. \ref{fig:Angle performance} shows the received power of the spoofing signal for different angular positions of the attacker with respect to the victim radar, using our mmWave testbed (see Sec. \ref{sec:sdr}). 
When the attacker is within the field-of-view of the victim radar, the victim radar is reliably spoofed. However, as the attacker moves away from $\pm 10^{\circ}$, the spoofing attack fails since the attacker is in the side-lobe of the victim radar.

\subsection{Faking Locations of Existing Obstacles}

The goal of this attack is to fake the location of an obstacle so that the victim AV considers that obstacle is not in its region-of-interest (ROI) (see definition in Appendix. \ref{sec:softwaremodules}). 
For example, in a multi-lane scenario, the attacker can deviate the obstacle in front of the AV (e.g., the leading vehicle) to a position away from the current lane, which spoofs the AV into crashing onto the leading vehicle. 
To achieve such goal, we propose three location-faking attack strategies: \textit{Random Signal Attack}, \textit{Synchronous Attack}, and \textit{Asynchronous Attack}.

\vspace{3pt}
\subsubsection{{\textbf{Random Signal Attack}}}
Considering the victim AV uses an optimal estimator to estimate the direction of detected obstacles, the Cramer-Rao bound \cite{abdelaziz2016security} for such estimator is: 
\begin{equation}
\frac{1}{2}[\sum_{i=1}^{K}Re\{x^{\dagger}_{rx}[i]\frac{\partial a }{\partial \theta }R^{-\frac{1}{1}}_{y}[I-a(a^{\dagger}a)^{-1}a^{\dagger}]R^{-\frac{1}{2}}_{y}\frac{\partial a}{\partial \theta}x_{rx}[i]\}]^{-1}\!,
\end{equation}
where $R_{y}\!=\!\frac{1}{L}\sum_{i=1}^{K}E[H_{adv}[i]x_{adv}[i]x^{H}_{adv}[i]H^{H}_{adv}[i]] \!+\! \sigma^2_{noise}I$ is the attacker signal covariance matrix with noise. 
It has been proved that the attacker's waveform that maximizes the variance of the estimation is drawn from a zero-mean Gaussian distribution \cite{abdelaziz2016security}. 
Therefore, in this Random Signal Attack, we use our mmWave testbed (see Sec. \ref{sec:sdr}) as the attacker and let it transmit a Gaussian waveform with controlled power to influence the victim radar.
From experiments, we prove that the attacker is able to deviate the angle of the detected objects by $\pm 5^{\circ}$. 
This attack can cause severe consequences on the AV's security as the true object's position is erroneously estimated. Then the AV could make the decision that the object does not exist in its ROI or the obstacle is in the adjacent lane, thus categorizing the obstacle as rather safe instead of flagging it.

\paragraph*{\textbf{Drawback of Random Signal Attack}}
Since the attacker's waveform is Gaussian distributed, the angle deviation introduced by the attacker is random. As a result, the Random Signal Attack cannot cause guaranteed and controlled attacking consequence. Therefore, we next proceed to develop attack strategies to generate deviations with controlled angles.

\subsubsection{{\textbf{Synchronous Attack}}}
\label{sec:Synchronous Attack}
Two challenges need to be addressed to fake the obstacle's location in a controlled manner: First, the attacker need to spoof an obstacle at a controlled position, especially the angular position (since the distance position can be spoofed using the strategy developed in Sec.~\ref{sec:distancespoofing}), thus creating an illusion the obstacles position has changed.
Second, the attack need to simultaneously obliterate the reflection signal from the existing obstacle.

\paragraph*{\textbf{Challenge 1}}
We start with addressing the first challenge, i.e., spoofing the angular position of the obstacle. 
The key idea is to employ multiple attackers located at different angular positions, hoping the combined spoofing signals from those attackers can create an illusion that the new obstacle is located at an angular position in between those attackers.

However, the challenge is that existing mmWave radar can already distinguish the signals from different attackers.
As discussed in Sec. \ref{sec: radar sensing}, the AV mmWave radar distinguishes object in the range $r_{m}$, Doppler $v_{m}$, and angle $\theta$ domain. 
Fig. \ref{fig:NotSyncAngleAttack} shows an example, the two attackers $s1$ and $s2$ are at the same distance $40m$ from the mmWave radar but with different angles ($-10^{\circ}$ and $7^{\circ}$).
In the left Range-Doppler map of Fig. \ref{fig:NotSyncAngleAttack}, the spoofed obstacles due to $s1$ and $s2$ have the same range bin $m$ but appear in different velocity bins $n$ and $n'$ due to different phase offset, resulting in two obstacles at $(r_{m}^{(1)},v_{n}^{(1)})$ and $(r_{m}^{(2)},v_{n'}^{(2)})$, which fails the spoofing attack.

To address the challenge, we introduce the synchronous attack, where the attackers are perfectly synchronised (in frequency, time, and phase), resulting in a correlated signal that make the spoofed obstacles appears in the same Range-Doppler bin $(r_{m}^{(1,2)},v_{n}^{(1,2)})$. Here $(1,2)$ refers to the obstacle spoofed by attacker 1 and 2. 
The next question is: \textit{what is the angular position of this spoofed obstacle $(1,2)$? Can we control the angle?}
As discussed in Sec.~\ref{sec: radar sensing}, the subspace-based direction estimation algorithm calculates the direction of the obstacles by separating the signal subspace $S(\theta_1,\theta_2)$ from the noise subspace $Z$ and finding the directions that are orthogonal to the noise subspace. 
The signal subspace $S(\theta_1,\theta_2)$ itself is a linear combination of the directions of all the received signals. 
We utilize such property and derive the following proposition for $K$ synchronous signals.

\begin{prop}\label{prop:synchronoussignals}
\textit{\textbf{For $K$ synchronous signals in the same Range-Doppler bin $\mathbf{(r_{m}^{1,\cdots,K},v_{n}^{1,\cdots,K})}$ with directions $\mathbf{\theta_1,\theta_2,\cdots,\theta_K}$, the spoofed obstacles direction $\mathbf{\theta_{spoof}}$ is the direction that maximizes $\mathbf{||S(\theta_1,\theta_2,\cdots,\theta_K)^Ha(\theta_{spoof})||^2}$ and is given by $\mathbf{\theta_{spoof}=mean(\theta_1,\theta_2,\cdots,\theta_K)}$}}.
\end{prop}

The synchronous attackers can launch the attack based on proposition 1. As a result, the victim AV identifies the combined spoofing signals from the multiple attackers as a single obstacle at $\mathbf{(r_{m}^{1,\cdots,K},v_{n}^{1,\cdots,K})}$ and direction $\theta_{spoof}$. The position of the spoofed obstacle is given by $(x_{spoof}(t),y_{spoof}(t))=\{x_{victim}(t)+d_{spoof}*cos(\theta_{spoof}),y_{victim}(t)+d_{spoof}*sin(\theta_{spoof})\}$, where $d_{spoof}$ is the spoofed distance corresponding to $(r_{m}^{(1,2)},v_{n}^{(1,2)})$ and $(x_{victim},y_{victim})$ is the position of the victim AV.

\vspace{3pt}
\paragraph*{\textbf{Challenge 2}}
Then we address the second challenge, i.e., \textit{How to obliterate the existing obstacle?} 
We employ an additional attacker to perform a jamming attack to overwhelm the genuine radar reflected signal.
The power of the jamming signal $P_{j}$ need to be higher than that of the reflected signal $P_r$ received by the victim AV. 
$P_{j}=\frac{P_{attacker}G_{attacker}\lambda^2}{(4\pi d_{attacker})^2}$, where $P_{attacker}$, $G_{attacker}$, and $d_{attacker}$ are the transmit power, antenna gain, and distance of the attacker.

\paragraph*{\textbf{Attacking Results}}
Fig. \ref{fig:SyncAngleAttack} shows the results of the synchronous attack. The scenario is the same as Fig. \ref{fig:NotSyncAngleAttack}. Now the two attackers $s1$ and $s2$ at directions $-10^{\circ}$ and $+7^{\circ}$ are \textit{synchronized}.
The Range-Doppler map shows that signals from the two attackers fall into the same bin. The point cloud shows that the attackers' spoofed obstacle is identified as a single obstacle $s1,s2$ with direction $4^{\circ}$ to the mmWave radar. 

\paragraph*{\textbf{Drawback of Synchronous Attack}}
The attackers need to be perfectly synchronized in frequency, phase, and transmission starting time. Any mismatch in frequency or phase of the waveform between the attackers, or the transmission starting time could invalidate the spoofing waveform in frequency and velocity.
For example, for a waveform with 300 MHz sweep bandwidth and 30 us sweep time, the frequency bin width in the range profile is $\frac{f_{sampling}*c}{2*slope*N_{fft}}=681.8 KHz$. If the carrier frequency offset (CFO) between the attackers is more than 681.8 KHz, the attack fails. Considering the mmWave radar works at 60 GHz or 77 GHz band, it is very difficult, if not impossible, to meet such synchronization requirement.

\vspace{3pt}
\subsubsection{{\textbf{Asynchronous Attack}}}
\label{sec: asyncsngleattack}

To avoid the stringent requirement of the synchronous attack, we propose a novel signal injection attack strategy that achieves the same performance only based on asynchronous attackers.
This attack is designed based on the following three observations:
\begin{itemize}
    \item The angle-of-arrival estimation method (see Sec. \ref{sec: angle estimation}) separates the signal subspace from the noise subspace. Any correlation in the subspace results in incorrect angle estimation, as described in Proposition \ref{prop:synchronoussignals}. 
    \item The direction estimation itself is agnostic to the baseband representation of the transmitted waveform. 
    \item The mmWave radar mixes the received waveform with the transmitted waveform to obtain the mixed signal in Eq. \ref{eq:fb}, which has a constant frequency of $2\pi\frac{B}{T}\tau$.
    Most signals other than a replica of the transmitted waveform (Eq. \ref{eq:xt}) result in mixed signals with high frequency that are filtered out by the low-pass filter of the victim AV's radar. However, if we use a filtered noise (correlated noise) to mix with the transmitted signal, a filtered signal can pass the low-pass filter as the output. 
\end{itemize}

The key idea of the proposed signal injection attack is to use filtered noise to introduce undesired correlation in the received radar signal samples, so that we can use multiple attackers to fake the location of existing obstacles.

\paragraph*{\textbf{Attack Procedure:}}
Without loss of generality, we consider there are two attackers in the following explanation. The $1$st attacker denoted as "attacker i" in Fig. \ref{fig:systemoverview} transmits a waveform with correlated noise $F(noise)$ and with delay $\tau_{delay}$ to spoof an obstacle at distance $d_{spoof}$. The transmitted waveform is given by 
\begin{equation}
\label{eq:attacker1signal}
TX_{spoof,1} \!=\! \sqrt{P_1}\left(\cos\left(2\pi f_{start} t' + \pi\frac{B}{T}t'^2\right) + F(noise)\right)\!,
\end{equation}
where $t'=(t-\tau_{delay})$, $P_1$ is the transmission power. The $2$nd attacker denoted as "attacker j" in Fig. \ref{fig:systemoverview} transmits a correlated filtered noise with power $P_2$:
\begin{equation}
\label{eq:attacker2signal}
TX_{spoof,2}=\sqrt{P_2}\cdot F(noise). 
\end{equation}

At the victim mmWave radar, after mixing with the radar waveform, the two spoofing signals become: 
\begin{align}
\label{eq:corrsigspoof}
&r_1=a(\theta_1)\frac{g_1}{PL(d_1)}\sqrt{P_1}\left(cos\left(\Phi\right) + F(noise)\right);\\
\label{eq:colorednoisespoof}
&r_2=a(\theta_2)\frac{g_2}{PL(d_2)}\sqrt{P_2}\cdot F(noise),
\end{align}
where $a(\theta_1), a(\theta_2)$ are the direction vectors corresponding to the directions of two attackers; $PL(d_1), PL(d_2)$ are the two distance dependent path loss; $\Phi=2\pi f_{start}\tau-\pi \frac{B}{T}\tau^2-2\pi\frac{B}{T}\tau t$.

Since the two spoofing signal $r_1$ and $r_2$ are correlated, the estimated direction $\theta_{spoof}$ of the spoofed obstacle is the direction that maximizes $||S(\theta_1,\theta_2)^H a(\theta_{spoof})||^2$.
Attacker i's waveform determines the distance of the spoofed obstacle. Attacker j's waveform together with attacker i influence the angle of the spoofed obstacle. Meanwhile, attacker j also performs as a jammer to obliterate the existing obstacle.
By this way, the two attackers do not need any synchronization but can simultaneously address the two challenges of the faking location attack described in Sec.~\ref{sec:Synchronous Attack}.

\begin{figure}
    \centering
    \includegraphics[trim={1cm 9cm 13cm 2cm},clip,width=1\columnwidth]{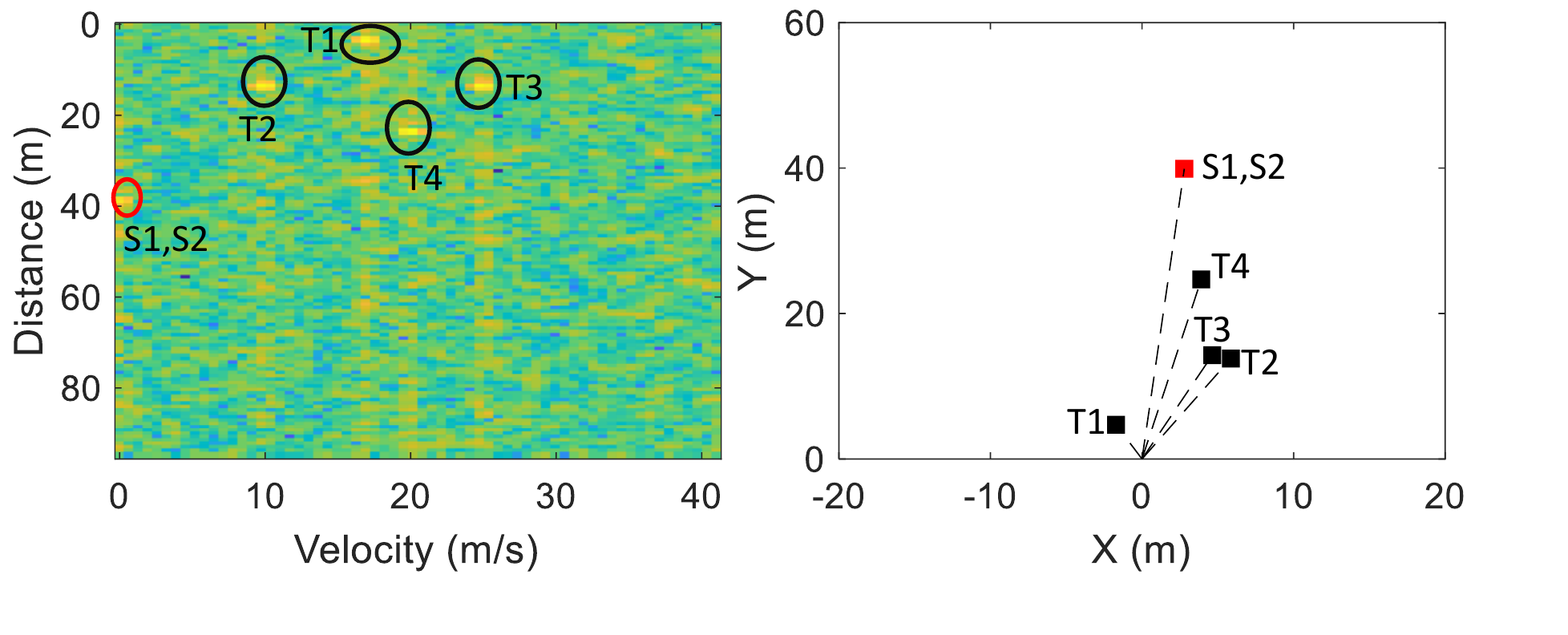}
    \vspace{-15pt}
    \caption{Range-Doppler map and point cloud showing the identified obstacles and the spoofed obstacle $s1,s2$. The asynchronous attackers perform distributed angle attack leading to the two spoofing signals resolved as single obstacle $s1,s2$.}
    \label{fig:CorrAngleAttack}
    \vspace{3pt}
\end{figure}

\paragraph*{\textbf{Attacking Results}}
Fig. \ref{fig:CorrAngleAttack} shows the results of two attackers $s1$ and $s2$ performing \textit{asynchronous} attack. 
The attackers spoofed obstacle is identified as a single obstacle $s1,s2$ at distance $40m$ and at angle $4^{\circ}$ to the mmWave radar.

\begin{figure}
    \centering
    \includegraphics[trim={1cm 10cm 21cm 1.5cm},clip,width=0.6\columnwidth]{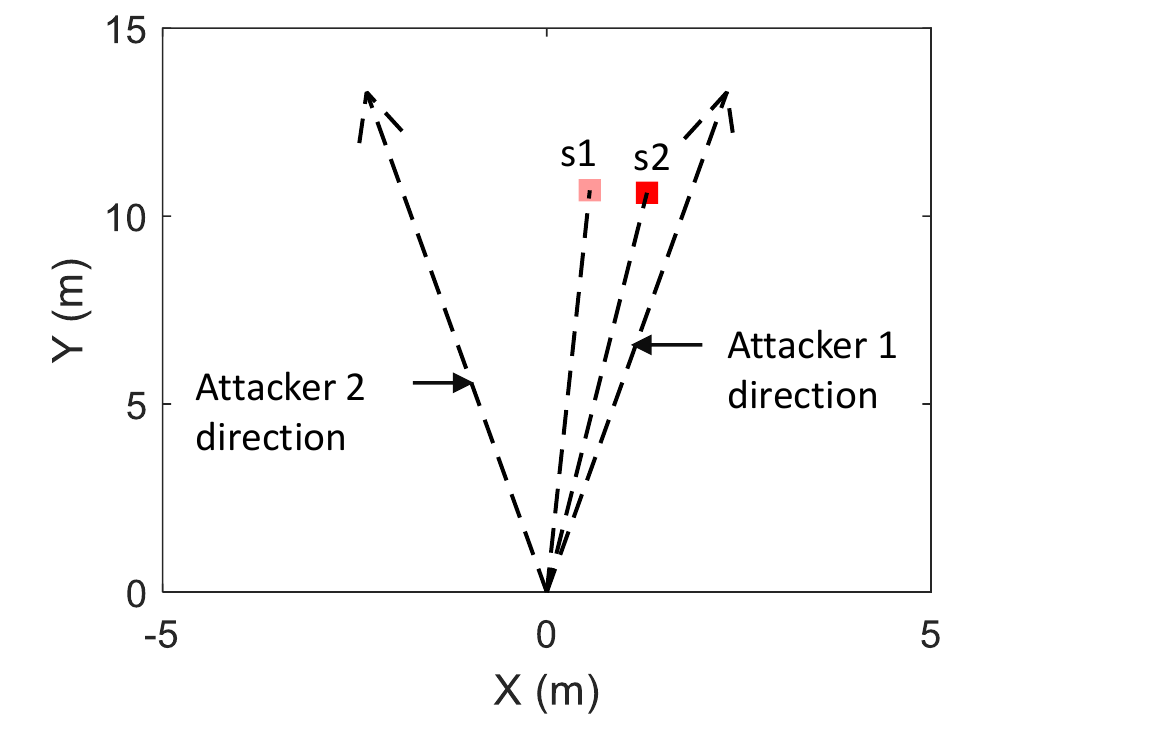}
    \caption{By varying transmission power, the asynchronous attackers can control the angular position of the spoofed obstacle.}
    \vspace{6pt}
    \label{fig:AngleAttackPower}
\end{figure}
We further perform experiments to show how the attackers control the position of the spoofed obstacle. In this experiment, the two attackers are located at a direction of $-10^{\circ}$ and $+10^{\circ}$, respectively, as shown in Fig. \ref{fig:AngleAttackPower}. 
Fig. \ref{fig:AngleAttackPower} shows the spoofed obstacle $s1$ at $+3^{\circ}$ with respect to the victim AV. By increasing the transmit power of attacker 1, we can move the spoofed obstacle's position to $s2$, which is $+7^{\circ}$ with respect to the victim AV. In Sec. \ref{sec:case study}, through the case studies of different driving scenarios, we show how such distributed and asynchronous attack could potentially trick the AV into making dangerous driving decisions.

\section{System Design}
To cause meaningful security consequences, the attacker need to continuously spoof the victim AV based on the combination of the two attack strategies discussed in Sec.~\ref{sec:attackdesign}. Moreover, beside launching the attacks, the attackers also need to: (1) sense the signal from the victim AV; and (2) track the position of the victim AV to continuously spoof it.

In this section, we introduce our attacker system based on the state-of-the-art mmWave testbed. We first present the mmWave testbed that lays the foundation of our attacker system. Then we discuss the main components of the attacker system. Finally we discuss how the attacker system performs continuous spoofing attack based on the mmWave testbed and the spoofing strategies developed in Sec.~\ref{sec:attackdesign}.

\vspace{-9pt}
\subsection{mmWave Radar Testbed}
\label{sec:sdr}
\vspace{-3pt}

\begin{figure}
    \centering
    \includegraphics[width=0.3\textwidth]{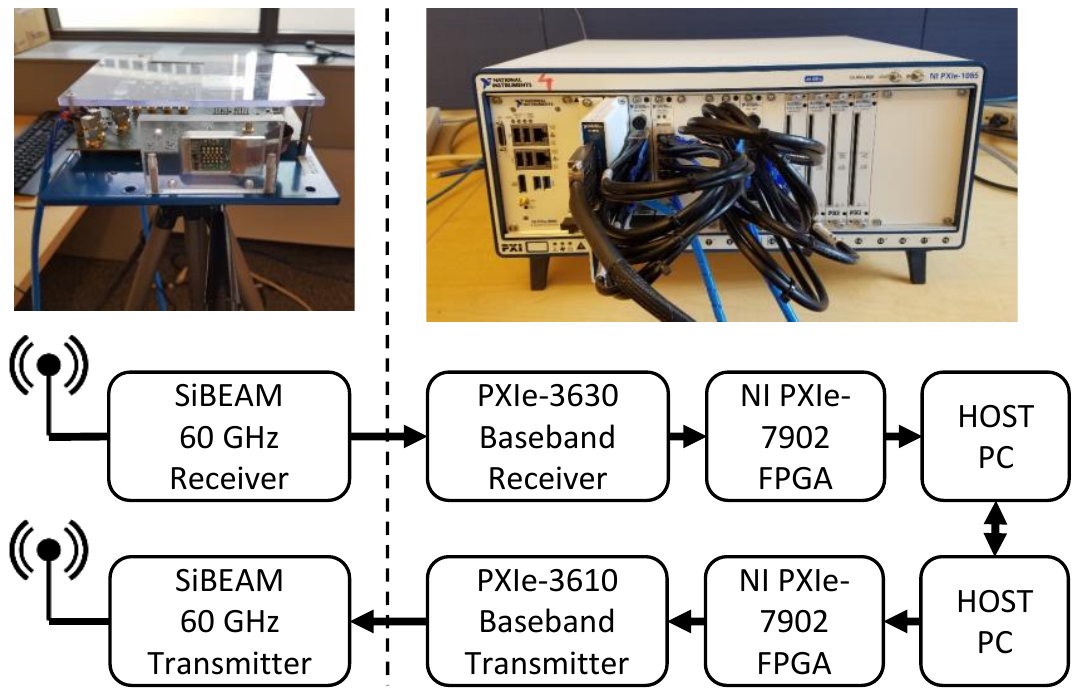}
    \vspace{3pt}
    \caption{The mmwave software defined radio testbed.}
    \vspace{9pt}
    \label{fig:X60}
\end{figure}

Our attacker testbed is based on a state-of-the-art software defined radio (SDR) mmWave transceiver system from NI \cite{niSDR}. The testbed architecture is shown in Fig. \ref{fig:X60}. Each mmWave SDR can be configured either as a transmitter or a receiver. On the transmitter side, each mmWave SDR has a PXIe-3610 DAC digital-to-analog converter. The PXIe-3610 is a 14-bit DAC with 3.072 GS/s sampling rate. It provides interface to analog baseband signal with a maximum bandwidth of 2 GHz. Each receive mmWave SDR has a PXIe-3630 ADC with 12 bits resolution and supports 3.072 GS/s sampling rate. 
The antenna module is a SiBEAM 60 GHz phased array that supports 2 frequency bands (60.48 GHz and 62.64 GHz center frequency). The mmWave SDR has a PXIe-7902 FPGA module to support physical layer baseband signal processing design.

\vspace{-9pt}
\subsection{Attacker System}
\label{sec:attacker system}
\vspace{-3pt}

\begin{figure}
    \centering
     \includegraphics[width=0.45\textwidth]{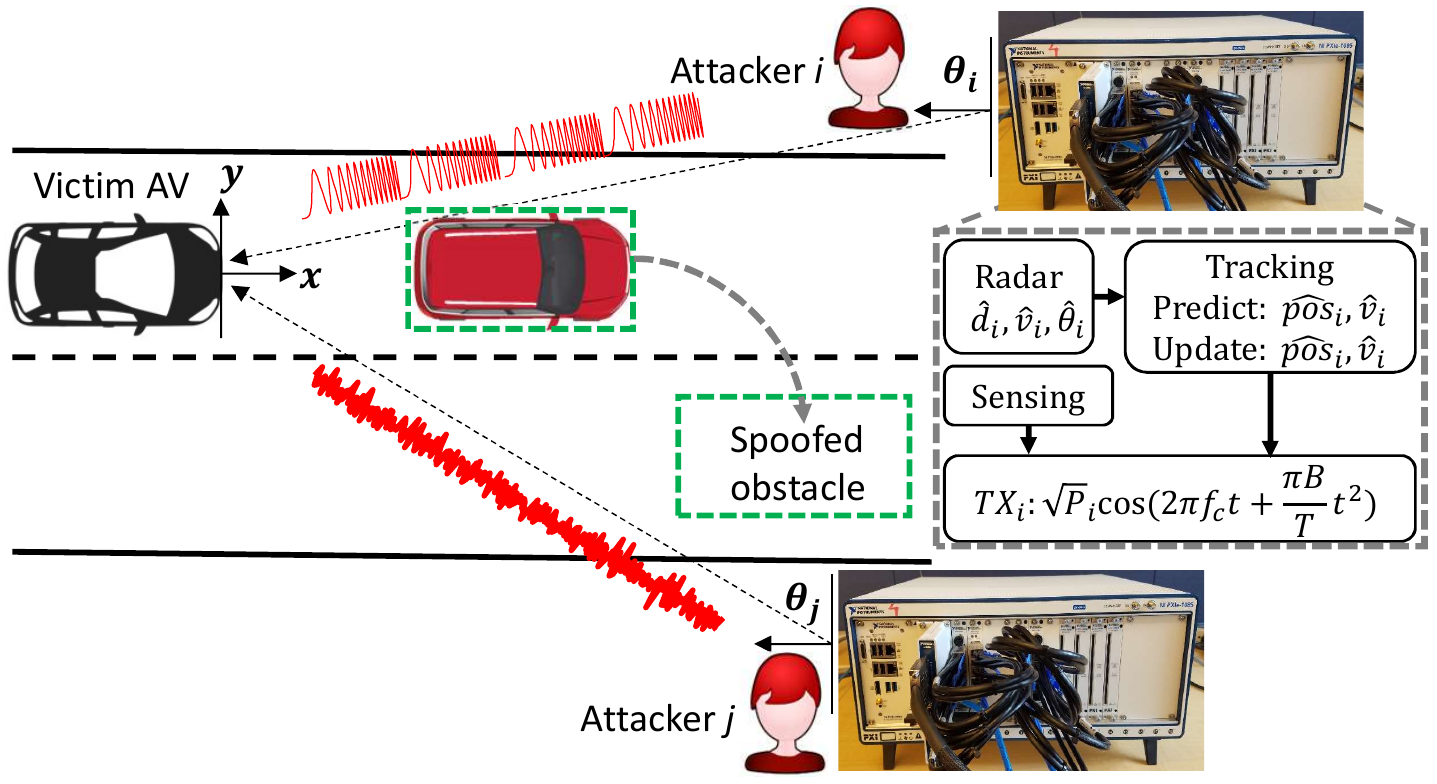}
    \caption{Conceptual AV spoofing scenario showing distributed attackers and the attacker system modules.}
    \vspace{6pt}
    \label{fig:attackScenario}
\end{figure}

As shown in Fig. \ref{fig:attackScenario}, each of the attacker systems consists of (1) \textbf{tracking system}, (2) \textbf{sensing system}, and (3) \textbf{transmitter}, working in tandem to form the attacker spoofing system.

\subsubsection{{\textbf{tracking system}}}
\label{sec:trackingsys}
\begin{figure}[htbp]
    \centering
     \includegraphics[trim = 4cm 9cm 4cm 8cm,clip,width=0.23\textwidth]{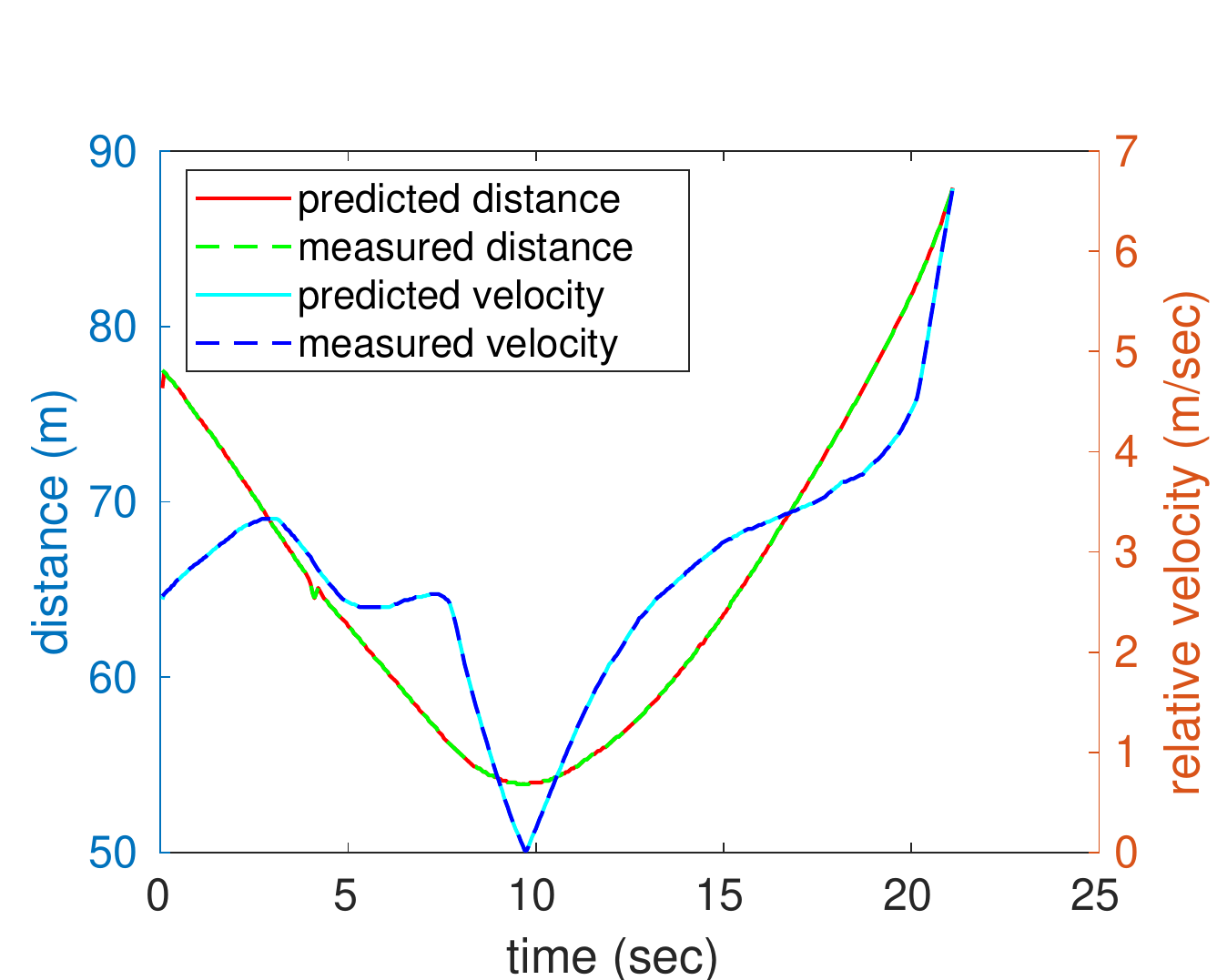}
    \includegraphics[trim = 4cm 9cm 4cm 8cm,clip,width=0.23\textwidth]{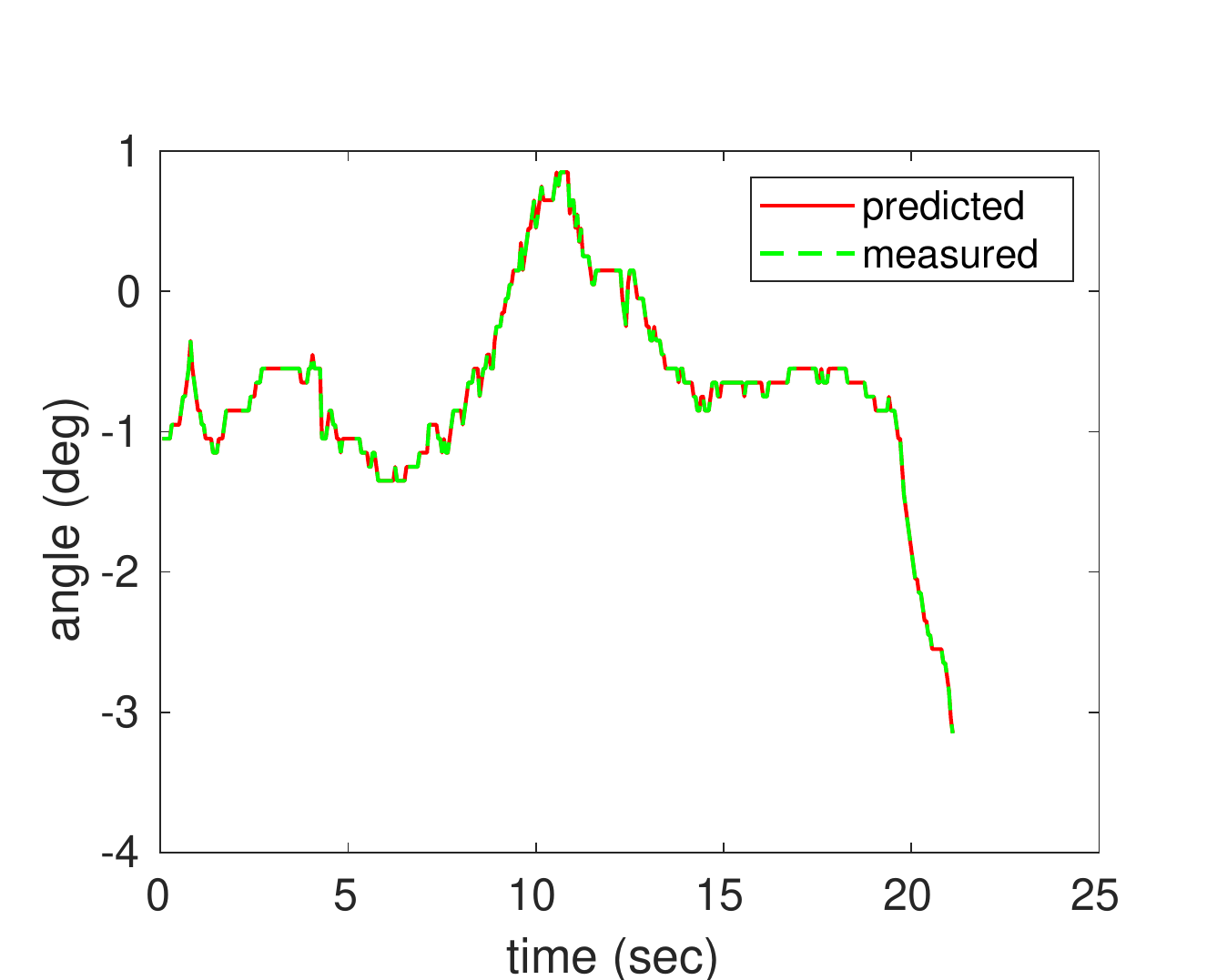}
    \caption{Tracking the distance,velocity, and angle of the victim AV using the Kalman filter.}
    \vspace{6pt}
    \label{fig:KF}
\end{figure}

The tracking system consists of radar and tracking modules. The radar is used to estimate the position and velocity of victim AV, based on Kalman filter (see Appendix. \ref{sec:KF}). To avoid interfering with the spoofing transmitter, the radar in tracking system operates in a different frequency band. This radar gives estimation results with a cycle period of $4.45 ms$. 
Fig. \ref{fig:KF} shows the distance, velocity, and angle predicted by the tracking system compared with the ground truth, which prove that the tracking system achieves accurate prediction of the victim's position and velocity.

\vspace{-6pt}

\subsubsection{{\textbf{sensing system}}}
\label{sec:sensingsys}
\begin{figure}[htbp]
    \centering
    \includegraphics[width=1\columnwidth]{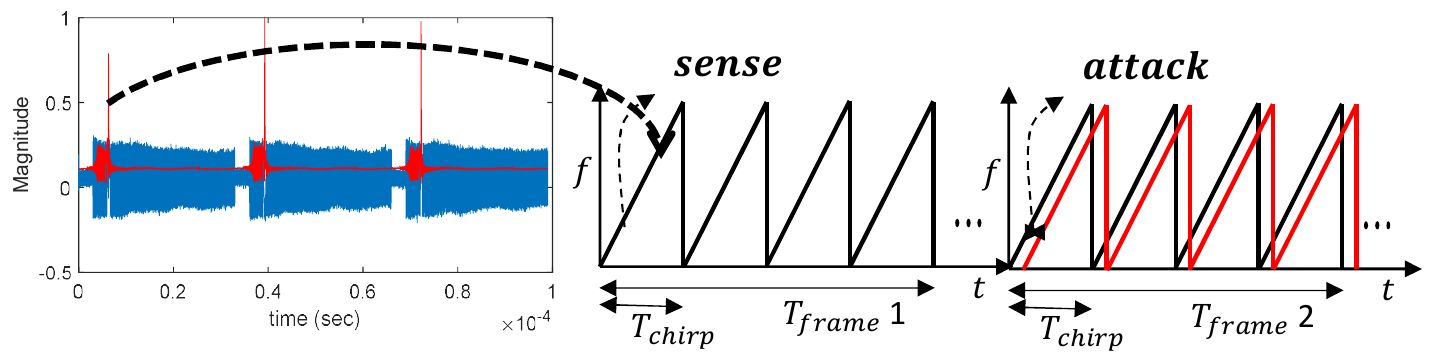}
    \caption{Detected mmWave signal from the victim AV.}
    \vspace{6pt}
    \label{fig:earlyDetect}
\end{figure}
The attacker needs to sense/detect the beginning of the transmitted mmWave signal (Eq. \ref{eq:xt}) from the victim AV. We use the receiver of our mmWave SDR testbed as the sensing system. 
According to the threat model, the attacker knows the frequency and waveform used by the victim mmWave radar (obtained from the radar's data sheet). To detect the signal from the victim AV, we use a correlator with the known victim chirp as a template \cite{ranganathan2012physical}. For a $33 us$ chirp and with the receiver sampling rate of 3.072 GSamples/Sec, the number of digitized samples is 101,376. Through experiments, we prove that the attacker only needs 5000 samples of the stored template chirp to reliably detect the mmWave signal from the victim AV. Fig. \ref{fig:earlyDetect} shows the received mmWave signal from the victim AV and the detected beginning of the signal using the correlator. With less than $5\%$ of the samples, the signal from the victim AV is detected with high accuracy.
 
\subsubsection{{\textbf{transmitter}}}
\label{sec:transmittingsys}
After detecting the signal from victim AV, the attacker's transmitter continuously spoofs the victim AV with the carefully designed signals described in Sec.~\ref{sec:attackdesign}. The spoofing waveform $x_{spoof}(t)$ is given by (1) Eq. \ref{eq:xt} to add fake obstacles, and (2) Eq. \ref{eq:attacker1signal} and Eq. \ref{eq:attacker2signal} to fake the locations of existing obstacles.
Due to the inherent delay between sensing the victim's signal and triggering the transmitter (limited by our current testbed), the spoof signals are not sent in the same time period as the signal sensing. Instead, in our attacker system, we sense the victim AV in a time period $t_1$ and trigger the spoofing in a different time period $t_2$ as shown in Fig. \ref{fig:earlyDetect}. 

\vspace{-9pt}
\subsection{Procedure of the Continuous Spoofing Attack}
\vspace{-3pt}
With the attacker system and its core modules described in Sec. \ref{sec:attacker system}, we can explain the attack procedure.
To reliably spoof an obstacle at a distance $d_{spoof}$, the adversary has to transmit the spoofing waveform under the delay constraints:
\begin{equation}\label{eq:distanceconstraint}
    t_{AV-Attacker}+t_{sensing}+t_{delay}+t_{Attacker-AV} = \frac{2*d_{spoof}}{c},
\end{equation}
where $t_{AV-Attacker}$ is the propagation delay from the victim AV to the attacker; $t_{sensing}$ is the time to detect the victim waveform; $t_{delay}$ is the delay after which the attacker transmits the spoofing waveform; and $t_{Attacker-AV}$ is the propagation delay from the attacker to the AV. Based on Eq. \ref{eq:distanceconstraint}, the attacker's distance $d_{Attacker}$ from victim AV is bounded by
\begin{equation}
    d_{Attacker}\leq \frac{\left(\frac{2*d_{spoof}}{c}-t_{sensing}-t_{delay}*c\right)}{2}.
\end{equation}

The limiting factors for attacker's position include sensing time $t_{sensing}$ and delay $t_{delay}$. The sensing system (Sec. \ref{sec:attacker system}) need anywhere between $1us$ to $3us$ on a state-of-the-art FPGA to sense the victim's signal. Assuming the attacker has a minimum switching speed of $10 ns$, and the objective of the attacker is to spoof an obstacle at 50 meters, the distance from the attacker to the victim AV should be $d_{Attacker} \leq 38.75 m$. That means the attacker is only be able to spoof obstacles at a distance further than its physical distance from the victim AV. In order to overcome this limitation, in our system, we delay the attack by several signal frames. As a result, the attacker can be further away than the spoofed obstacle.

\par The attacker system works as follows. The multiple attackers are coordinated (e.g., through WiFi) but asynchronous. At the beginning of the attack, i.e., time $t_0$, each of the attackers estimates the position of the victim AV using the tracking system (Sec. \ref{sec:trackingsys}), and continuously tracks the position of the victim AV for subsequent time intervals. At time $t_k$, the attackers detect the mmWave signal from the victim AV using the sensing system (Sec. \ref{sec:sensingsys}). Based on the estimated position of the victim AV and the attack goal (Sec. \ref{sec:case study}), the attackers transmit corresponding spoofing signals with delay $\tau_{delay}$. The two-way propagation time due to sensing and spoofing needs to be considered when determining $\tau_{delay}$. The attackers continuously calculate the spoofing delay $\tau_{delay}$ using the input from the tracking system. 
The spoofing strategy (transmitted waveform) depends on the goal of the attacker. For \textit{AV stalling attack} (Sec. \ref{sec:AVstall}), \textit{Hard Braking attack} (Sec. \ref{sec:hardbraking}), and \textit{Lane change attack} (Sec. \ref{sec:lanechange}), the attacker uses the strategy described in Sec. \ref{sec:distancespoofing}. For multiple stages attack scenarios  \textit{Multi-stage Attack} (Sec. \ref{sec:scenario4}) and \textit{Cruise control attack} (Sec. \ref{sec:tesla}), multiple distributed attackers employ combination of strategies described in Sec. \ref{sec:distancespoofing} and Sec. \ref{sec: asyncsngleattack}.

\section{Field Experiments and Security Analysis}
\label{sec:Security Analysis}
\vspace{3pt}

\begin{figure*}
    \centering
    \begin{minipage}{0.44\textwidth}
    \centering
    \includegraphics[width=0.9\textwidth]{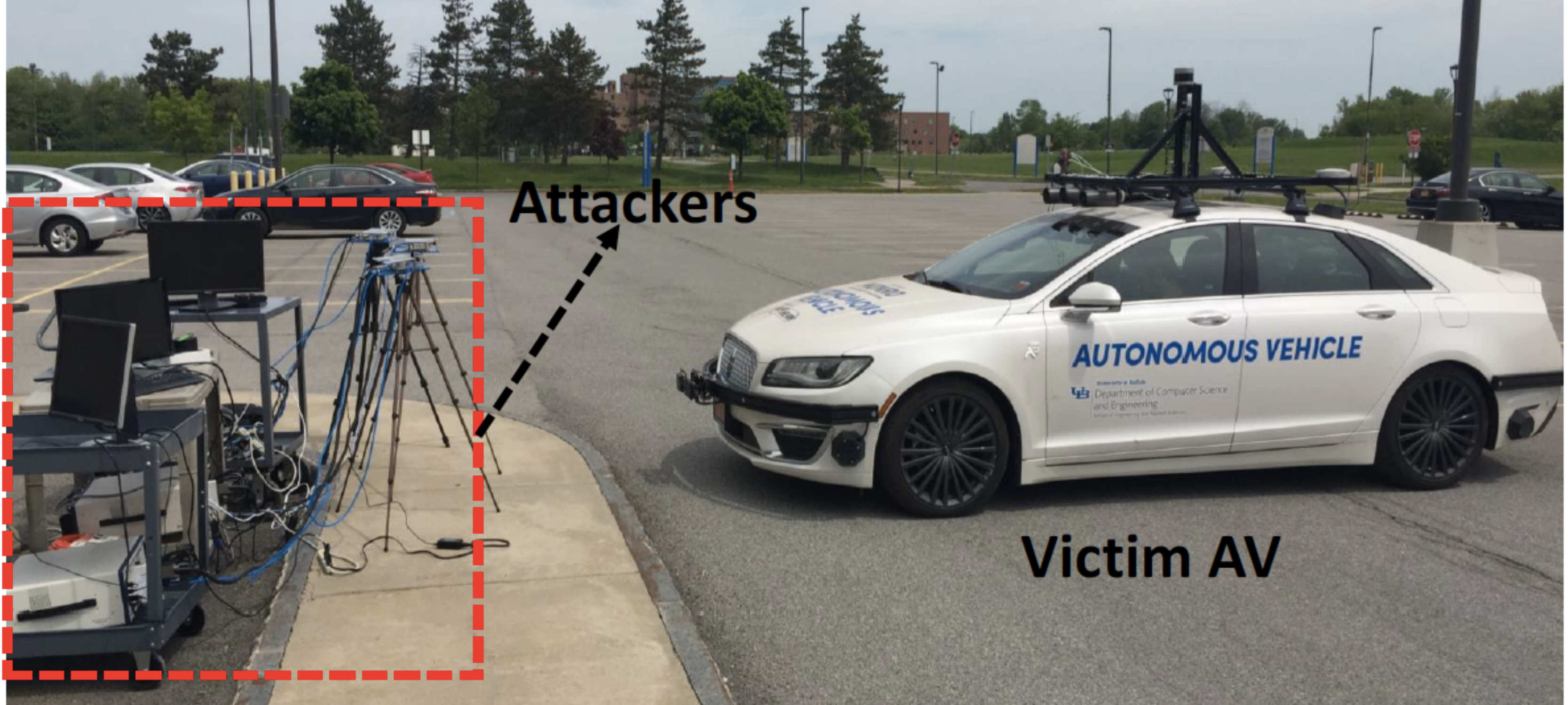}
    \vspace{3pt}
    \caption{Experiment set up showing the attacker mmWave testbed and the Lincoln MKZ-based AV testbed at University at Buffalo.}
    \label{fig:exp_setup}
    \end{minipage}
    ~
    \begin{minipage}{0.25\textwidth}
        \includegraphics[trim={0cm 7cm 19cm 1cm},clip,width=1\textwidth]{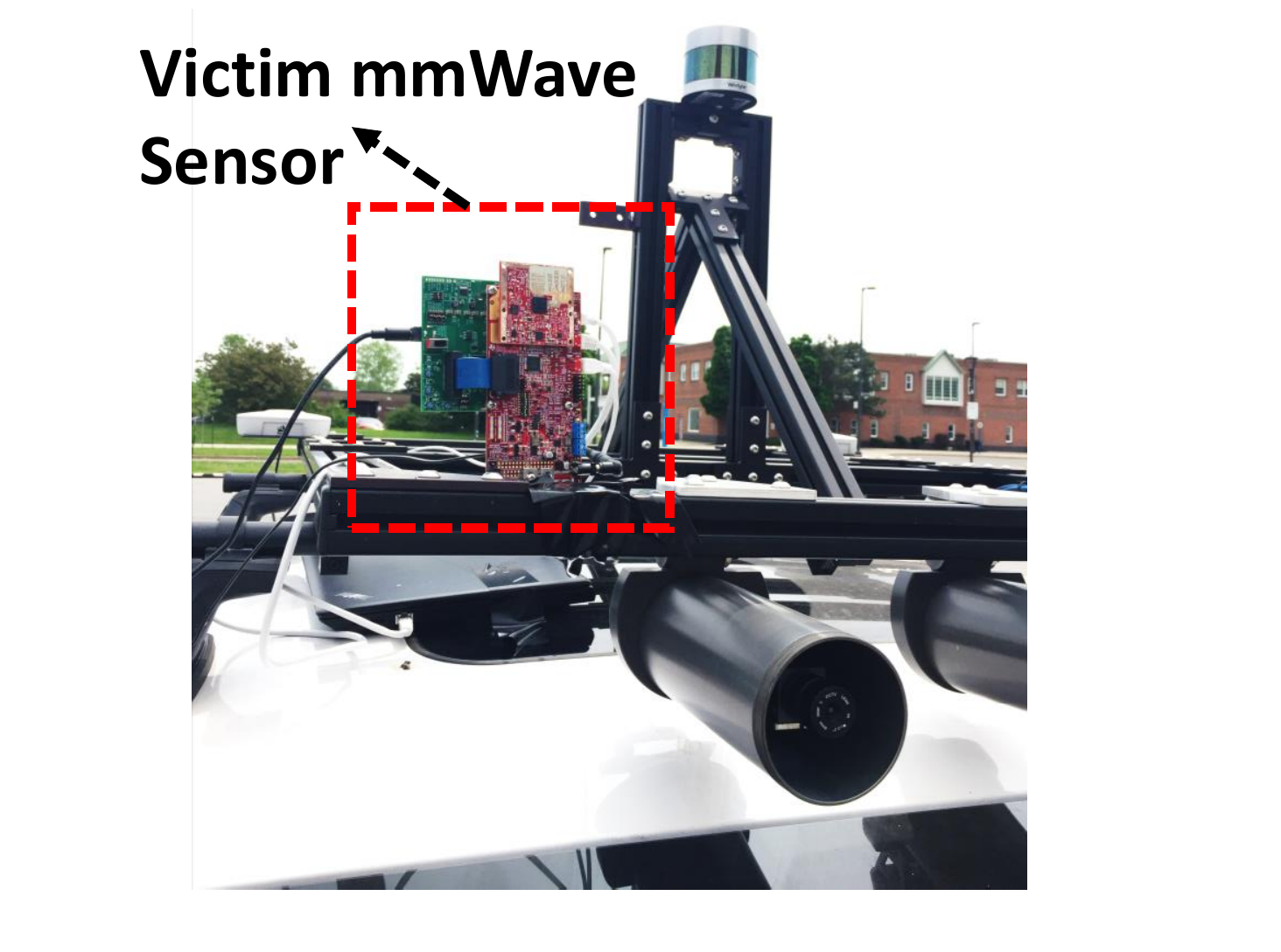}
    \caption{Experiment set up showing the radar mounted on the victim AV.}
    \label{fig:radar_mount}
    \end{minipage}
    ~~~
    \begin{minipage}{0.25\textwidth}
    \scalebox{0.8}{
    \begin{tabular}{|c|c|}
    \hline
         Parameter & Value \\
    \hline
         Sweep bandwidth & 300 MHz \\
    \hline
        Ramp slope & 10 MHz/us \\
    \hline
        Inter-chirp duration & 3 us \\
    \hline
        Number of chirps & 128 \\
    \hline
        Chirp duration & 30 us \\
    \hline
        Total frame time & 4.224 ms\\
    \hline
        Active frame time & 3.84 ms\\
    \hline
        Start frequency & 62.61 GHz \\
    \hline
    \end{tabular}}
    \vspace{6pt}
    \captionof{table}{Long range radar parameters of the victim AV.}
    \label{table:radarparams}
    \end{minipage}
    \vspace{6pt}
\end{figure*}

\begin{figure*}
\vspace{-6pt}
\centering
\begin{minipage}[t]{0.28\textwidth}
    \centering
    \includegraphics[trim={1cm 10cm 23cm 1.5cm},clip,width=0.93\textwidth]{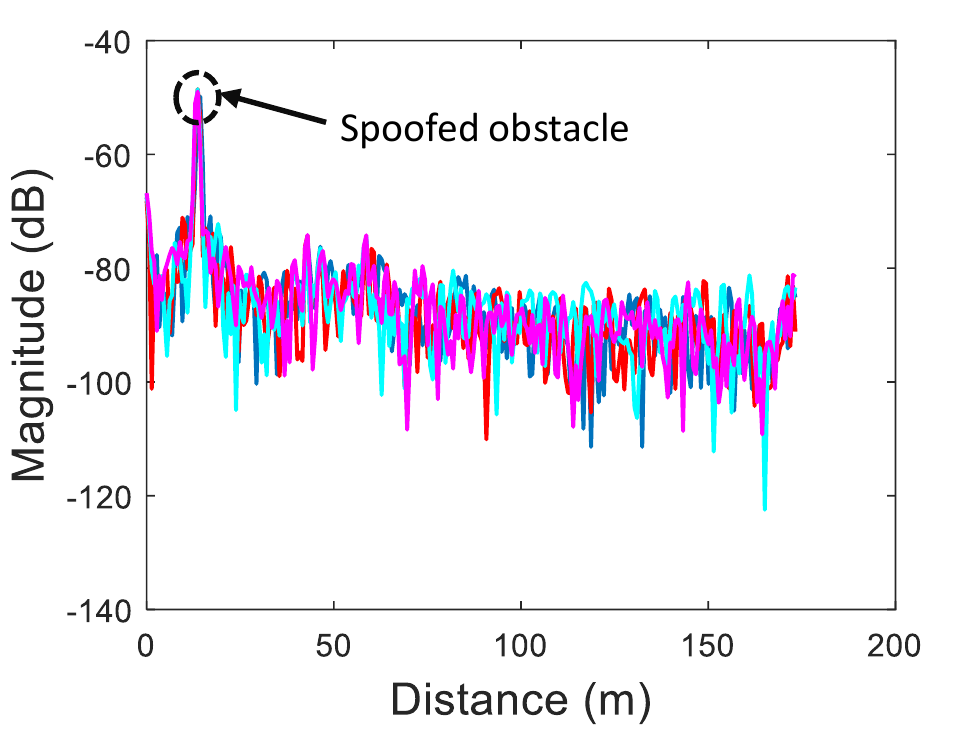}
    \vspace{-3pt}
    \caption{AV stalling attack: Spoofed obstacle at 13m at multiple frames overlayed.}
    \label{fig: AVStallAttackobstacle}
\end{minipage}
~
\begin{minipage}[t]{0.7\textwidth}
\centering
\includegraphics[trim={4cm 8cm 9cm 3cm},clip,width=0.65\textwidth]{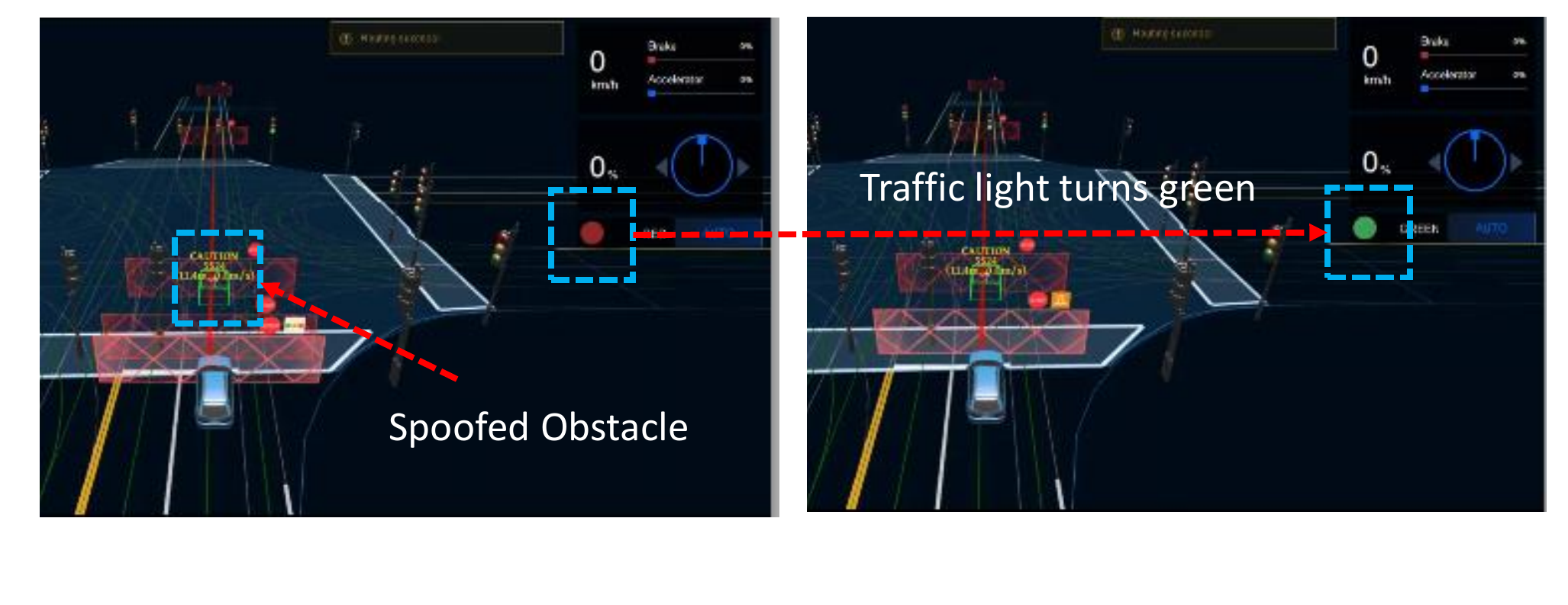}
\vspace{-6pt}
\caption{End-to-end demonstration of AV stalling attack. The AV is waiting at a red light traffic junction. The attacker spoofs an obstacle at $13m$ in front of the AV. The AV does not move ahead after the traffic light turns green, since it \textit{sees} a spoofed obstacle.}
\vspace{6pt}
\label{fig: AVStallAttack}
\end{minipage}
\end{figure*}

\begin{figure*}
\vspace{-6pt}
    \centering
    \begin{minipage}[t]{0.28\textwidth}
    \centering
    \includegraphics[trim={1cm 10cm 23cm 1cm},clip,width=0.9\textwidth]{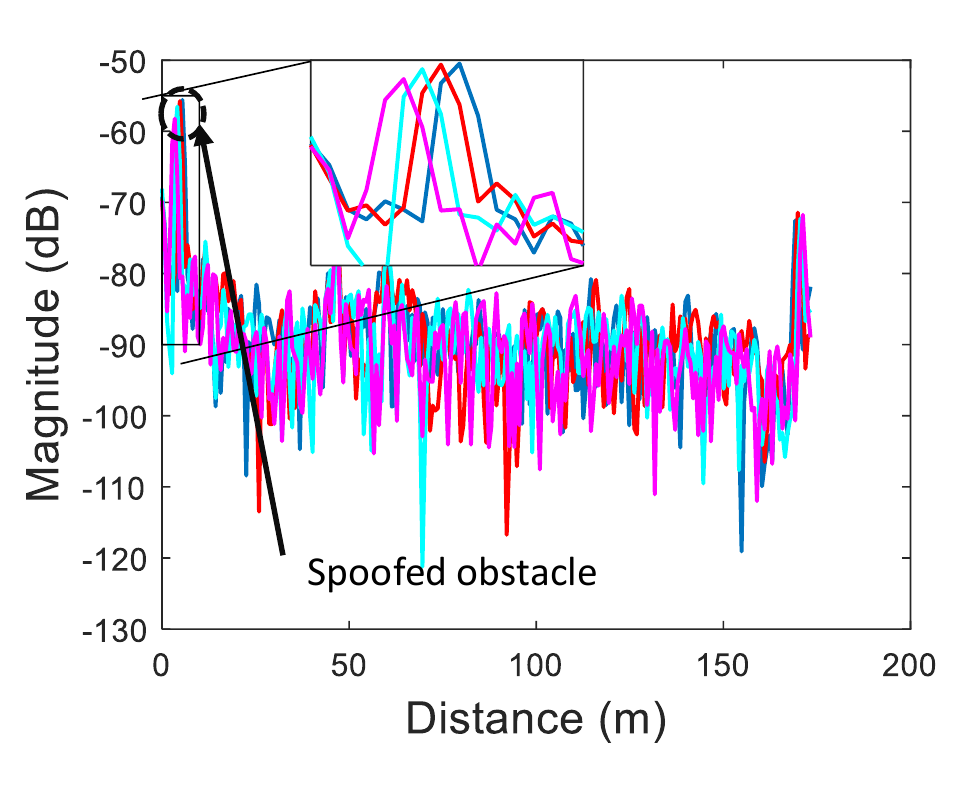}
    \vspace{-6pt}
    \caption{Hard braking attack: Spoofed obstacles at multiple time frames are overlayed.}
    \label{fig:AVhardbrakeobstacle}
    \end{minipage}
    ~
    \begin{minipage}[t]{0.7\textwidth}
    \centering
    \includegraphics[trim={5cm 9cm 13cm 3cm},clip,width=0.6\textwidth]{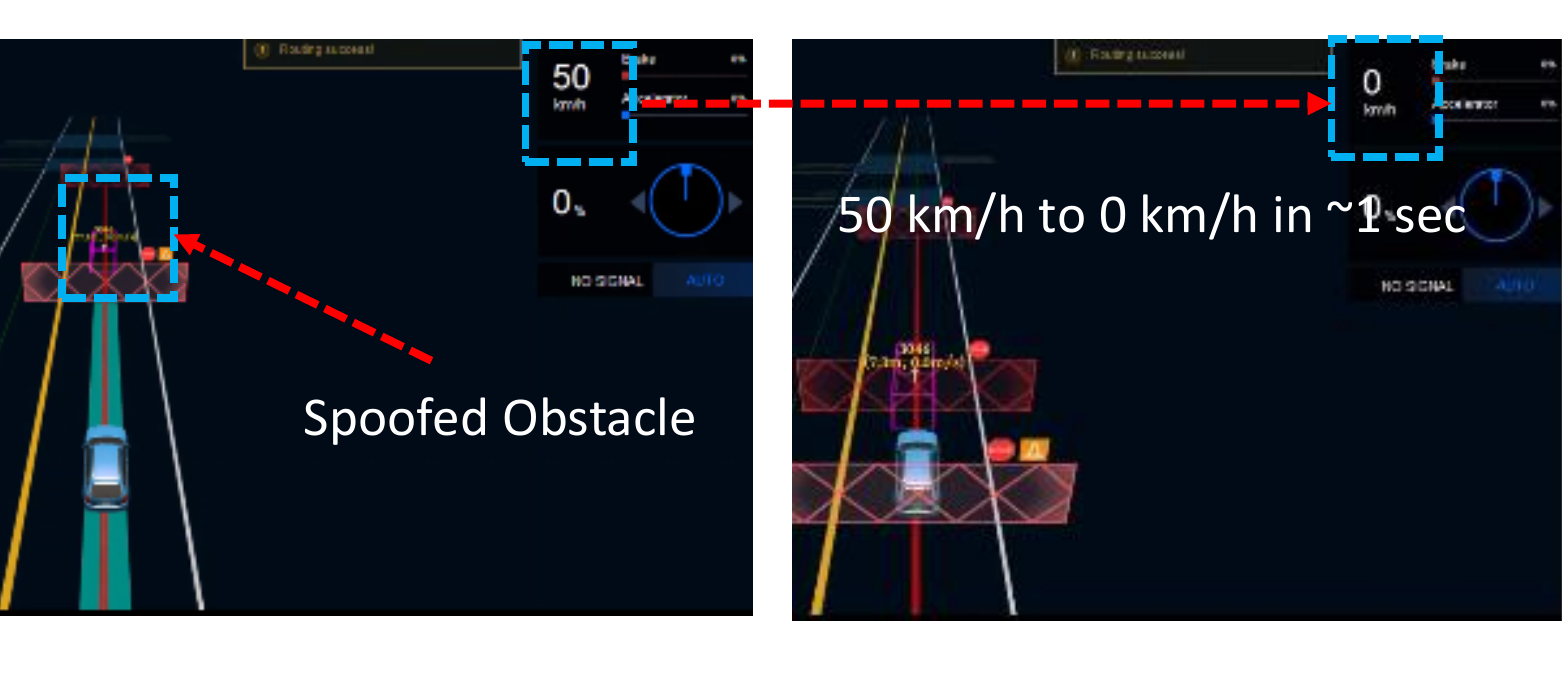}
    \vspace{-6pt}
    \caption{End-to-end demonstration of AV hard braking attack. The AV is travelling at a speed of 50 km/hr. The attacker spoofs an obstacle at $27m$ in front of the AV. The AVs perception module detects the sudden obstacle and makes an emergency braking decision.}
    \vspace{10pt}
    \label{fig:AVhardbrake}
    \end{minipage}
\end{figure*}

We construct five real-world driving scenarios to investigate the end-to-end security of AV under mmWave sensing spoofing attacks. 
We perform real-world experiments on a Lincoln MKZ autonomous vehicle testbed \cite{LincolnMKZ}, \cite{dmowski2019research}, which was developed at University at Buffalo, as shown in Fig. \ref{fig:exp_setup}. The AV is equipped with an TI IWR6843 radar \cite{TI6843} that uses the $60 GHz$ band for sensing. Table \ref{table:radarparams} gives the parameters of the radar. In addition to the mmWave radar, this AV testbed also has a LiDAR and several cameras. 
We use Baidu Apollo software \cite{ApolloBaidu} as victim AV's perception, planning, and navigation software. The parameters of the Baidu Apollo AV software are given in Table \ref{table:Baiduparam} in Appendix. \ref{sec:BaiduApolloParam}. We perform the \textit{software-in-the-loop} drive-by experiments using the Lincoln MKZ AV. We first spoof the Lincoln MKZ's mmWave radar. Then the detection results from the victim radar are input to the Baidu Apollo software to derive the decision of AV. This process is repeated for each of the time steps of the scenarios to evaluate the end-to-end security of the victim AV. Due to safety considerations, we performed our drive-by field experiments in an open parking lot. The AV is driven with a speed of approximately $10 mph$. Fig. \ref{fig:exp_setup} shows the experiment set up with the mmWave testbed used as attackers. According to the threat model, we keep the attackers (i.e. the mmWave testbed) stationary and placed them on the roadside. The mmWave radar is mounted on the roof of the AV as shown in Fig. \ref{fig:radar_mount} due to cabling limitations.

\label{sec:case study}

\subsubsection{{\textbf{Scenario 1: AV Stalling Attack}}}
\label{sec:AVstall}

In this attack, the goal of the attacker is to spoof an obstacle in front of a stationary AV to stall it. For example, an AV waiting for the red light in a traffic junction might not move ahead when such an obstacle is detected by its radar radar, causing traffic confusion. 

\par We performed experiments to emulate such a scenario. The victim Lincoln MKZ AV was at a distance of 20m from the attacker. After sensing and tracking the victim's signal as described in Sec. \ref{sec: radar sensing} at time $t_{k}$, the attacker transmits spoofing waveform in Eq. \ref{eq:xt} continuously to spoof an obstacle at $13m$ in front of the victim AV. The location of the spoofed obstacle is given by $\{x_{spoof}(t_k),y_{spoof}(t_k)\}=(x_{victim}(t_k)+d_{spoof}*cos(\theta))+\epsilon_{var},y_{victim}(t_k)+d_{spoof}*sin(\theta))+\epsilon_{var})$. Fig. \ref{fig: AVStallAttackobstacle} shows the spoofed obstacle at $13m$ for consecutive frames. Fig. \ref{fig: AVStallAttack} shows the Baidu Apollo view with input from the victim radar. When the spoofed obstacle is within $13m$ distance, the AV chooses not to move ahead even when the traffic light turns green. 
In Baidu Apollo, if the spoofed obstacle is further than $15m$ and if there is an adjacent lane (two-lane road), the AV performs a side-pass and drive around the obstacle. In such scenario, the attacker could utilize the multiple obstacle spoofing method described in Sec. \ref{sec:multiplespoof} to simultaneously spoof multiple obstacles on both the lanes.

\subsubsection{{\textbf{Scenario 2: Hard Braking Attack}}}
\label{sec:hardbraking}

In this attack, the objective of the attacker is to spoof an obstacle to force the victim AV make a hard braking decision thus endangering the safety of the occupants of the AV as well as other cars on the road. 
The attacker tracks the position $pos_{victim}$ of the victim AV as in Sec. \ref{sec:trackingsys} and transmits a spoofing signal to spoof an obstacle near the victim AV. In a multi-lane scenario, the attacker spoofs multiple obstacles so that the victim AVs planning and control module does not initiate a sudden lane change action. We spoofed an obstacle at $27m$, and the detected obstacle was fed to the AV software. To imitate an obstacle slowing down in front of the AV, the attacker gradually decreases the spoofed obstacle distance with respect to the victim AV. 
Fig. \ref{fig:AVhardbrakeobstacle} shows the spoofed obstacle for consecutive frames and Fig. \ref{fig:AVhardbrake} shows the corresponding scenario where the victim AV is travelling at a speed of 50 km/hr. The attacker spoofs an obstacle at $27m$ in front of the AV. At a speed of 50 km/hr, the AV would cover $27m$ in $1.944$ sec, which is less than the average driver reaction time of $2.3$ sec. The AV initiates a hard braking decision in this case which could potentially lead to fatal situation. 

\subsubsection{{\textbf{Scenario 3: Lane Change Attack}}}\label{sec:lanechange}


\begin{figure}
    \includegraphics[trim={1.5cm 13cm 21cm 2.5cm},clip,width=1\columnwidth]{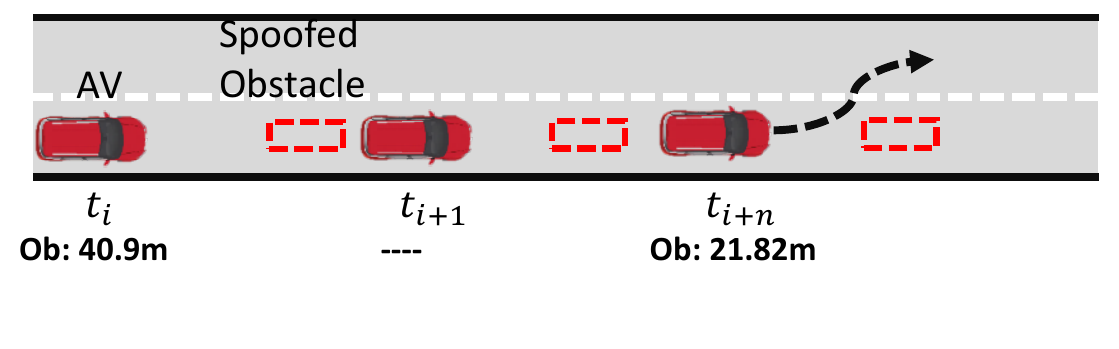}
    \vspace{-25pt}
    \caption{Illustration of the \textit{planning} module (lane change) attack scenario. Victim AV is travelling at right lane before it changes lane due to spoofed obstacle.}
    \label{fig:scenario3}
\end{figure}
\begin{figure}
    \centering
    \begin{minipage}[t]{0.5\columnwidth}
    \includegraphics[trim={0cm 10.5cm 24cm 1cm},clip,width=0.95\textwidth]{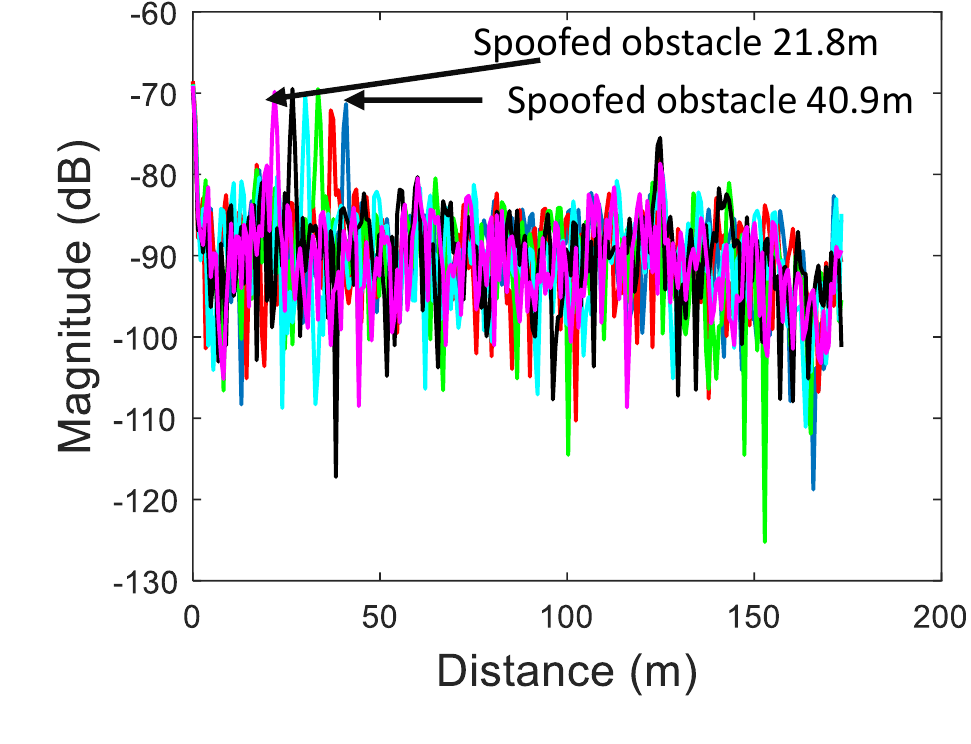}
    \caption{Lane change attack: Spoofed obstacles at multiple time frames are overlayed.}
    \label{fig:lanechangeobstacle}
    \end{minipage}
    ~
    \begin{minipage}[t]{0.41\columnwidth}
    \centering
    \includegraphics[trim={0cm 2cm 5cm 2cm},clip,width=0.9\textwidth]{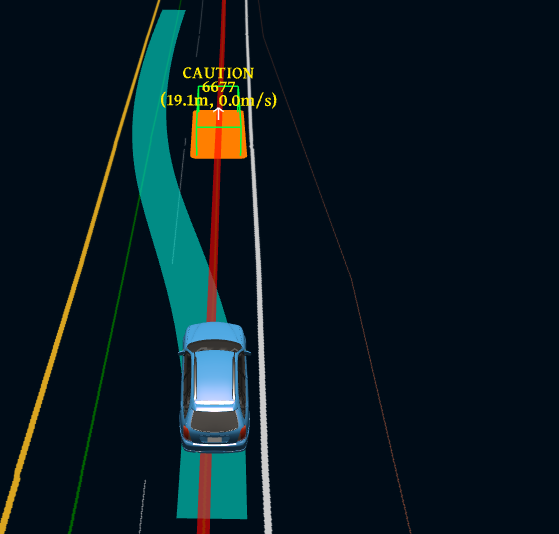}
    \vspace{3pt}
    \caption{The victim AV's planning algorithm charts an alternate path due to a spoofed obstacle.}
    \vspace{6pt}
    \label{fig:lanechange}
    \end{minipage}
\end{figure}

\par
The goal is to spoof the AV into making dangerous lane change decisions.
We construct a scenario shown in Fig. \ref{fig:scenario3}. For this attack experiment, the Lincoln AV is initially at a distance of $25m$ from the attacker. The attacker detects the mmWave signal from the victim AV and tracks the position of the victim AV as described in Sec. \ref{sec:sensingsys} and Sec. \ref{sec:trackingsys}, respectively. Upon estimating the position of the AV, the attacker spoofs an obstacle in the lane in which the victim AV is traveling. 
The attacker spoofs an obstacle at $~40m$ at time $t_i$. To imitate the distance decreasing effect, the attacker spoofs the obstacle with decreasing distance at subsequent time frames as shown in Fig. \ref{fig:lanechangeobstacle}. The victim AV was travelling at a speed of $30 mph$. Fig. \ref{fig:lanechange} shows the decision taken by the victim AV to change lane due to spoofed obstacle in its lane. As discussed in Appendix. \ref{sec:pathplanning}, the cost function of the planning module decides the appropriate path decision based on various factors. Eq. \ref{eq:costfunc} shows that when the spoofed obstacle greater than the collision distance threshold, the cost of the lane in which the AV is travelling increases compared to other lanes and the AVs planning module searches for other alternative path. The attacker could take advantage knowing the planning algorithm to deviate the victim AV from the planned path. Another observation we noticed is that, the AV decides to change lane only when it is within a threshold distance from the spoofed obstacle. Until it reaches the threshold distance, the AV slows down.

\subsubsection{{\textbf{Scenario 4: Multi-Stage  Attack}}}\label{sec:scenario4}

\begin{figure}[htb!]
    \centering
    \includegraphics[trim={1.5cm 8cm 19.5cm 2cm},clip,width=1\columnwidth]{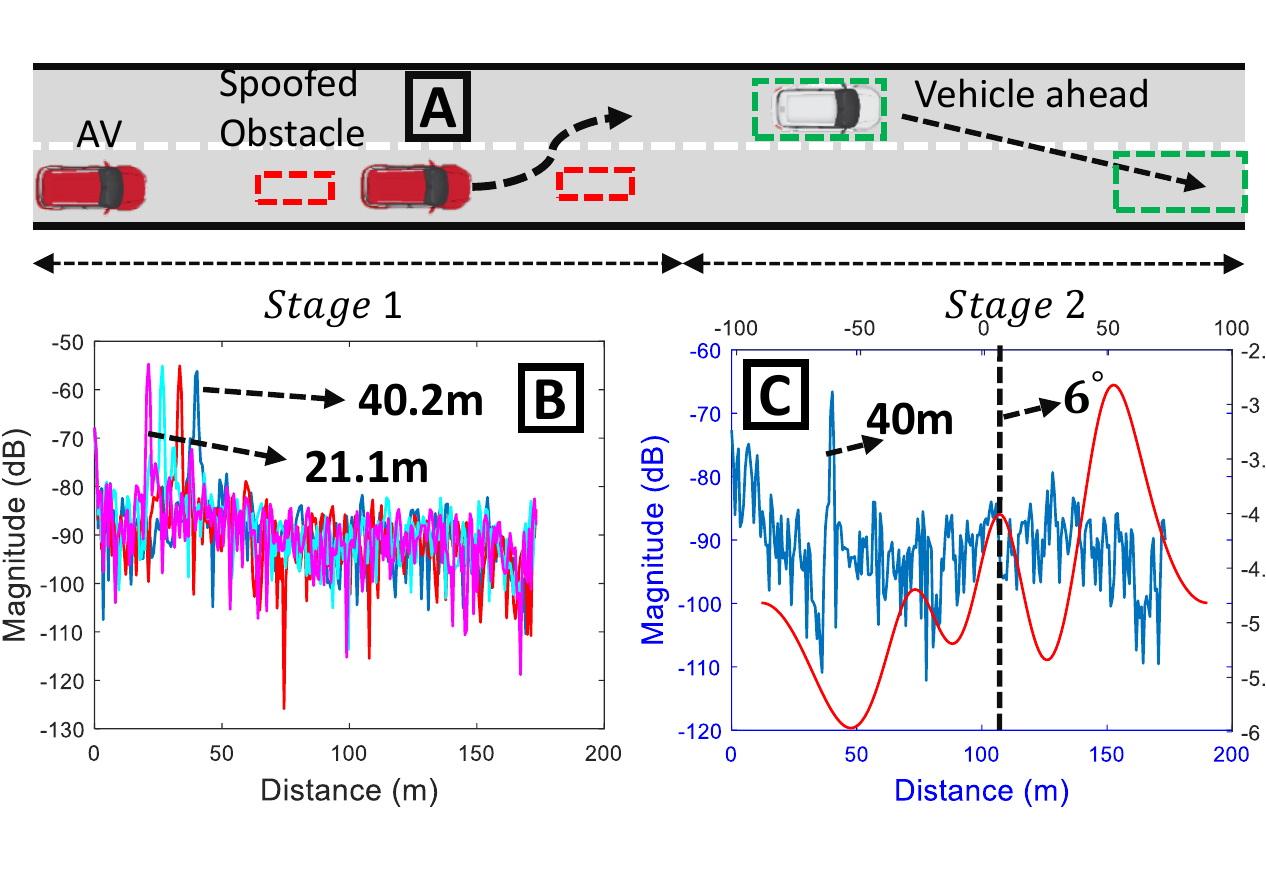}
    \vspace{-20pt}
    \caption{Illustration of scenario 4. At stage 1, the victim AV changes lane due to a spoofed obstacle. At stage 2, the vehicle ahead is spoofed to a different location.}
    \vspace{6pt}
    \label{fig:scenario4}
\end{figure}

We construct a multiple-stage attack scenario by combining the intuitions and AV driving behaviour from the previous attack scenarios, which could lead the victim AV to a serious accident. We emulate a complex AV driving scenario shown in block [A] in Fig. \ref{fig:scenario4}, where the victim AV is initially cruising on a multi-lane road. 
We use distributed attack strategy discussed in Sec. \ref{sec: asyncsngleattack}. The objective of the attacker is to force the victim AV to make a dangerous lane change (see Sec. \ref{sec:lanechange}) and subsequently crash on to a vehicle ahead of it.

As shown in Fig. \ref{fig:scenario4} [A], 
the victim AV is cruising in lane 1 while another vehicle travels in lane 2. To force a lane change, the attacker spoofs an obstacle at distance $d_{spoof}$ (40.2m) in front of the victim AV, as shown in block [B] in Fig. \ref{fig:scenario4}. The corresponding position of the spoofed obstacle at time $t_i$ is $(x_{spoof}(t_{i}),y_{spoof}(t_{i}))=\{x_{victim}(t_{i})+d_{spoof}*cos(\theta)+\epsilon_{var},y_{victim}(t_{i})+d_{spoof}*sin(\theta)+\epsilon_{var}\}$. In order to consistently keep the cost of the current lane higher and to force the AV to think, the vehicle ahead is slowing down, the attacker spoofs obstacles at decreasing distance from $40.2m$ to $21.2m$ at subsequent time intervals, as shown in block [B] in Fig. \ref{fig:scenario4}. The position of the spoofed obstacles in subsequent time intervals are $(x_{spoof}(t_{i+1}),y_{spoof}(t_{i+1}))=\{x_{victim}(t_{i+1})+(d_{spoof}-v_{rel}*c)*cos(\theta)+\epsilon_{var},y_{victim}(t_{i+1})+(d_{spoof}-v_{rel}*c)*sin(\theta)+\epsilon_{var}\}$, where $v_{rel}$ is the relative velocity between the victim and the adversary. 

At \textit{stage 2}, two attackers perform the obstacle-deviation attack (Sec. \ref{sec: asyncsngleattack}). The attackers spoof the location of the vehicle in lane 2 at $40m$ and at an angle of $6^{\circ}$ to the victim AV, as shown in block [C] in Fig. \ref{fig:scenario4}. Such attack makes the vehicle ahead of the victim AV appear in a different lane than its true lane, which leads the victim AC to a potentially fatal accident. As discussed in Sec. \ref{sec:hardbraking}, on the highway driving conditions, with a typical lane change speed of $45$ to $50$ mph, a lead vehicle at $30m$ that \textit{disappears} from the victim AVs mmWave radar view could turn in to a fatal accident. In addition, as we show in the next attack scenario, when the leading vehicle \textit{disappears} from the field-of-view, the victim AV could also accelerate to match the lane speed, causing a high speed crash.

\subsubsection{{\textbf{Scenario 5: Cruise Control Attack}}}\label{sec:tesla}

Traffic-Aware Cruise Control (TACC) systems can keep the AV cruising at a set speed while at the same time adjusting its speed according to the traffic. Although such functionality offers comfort to the drivers, 
a powerful attacker can take advantage of the TACC techniques to manipulate the driving behaviour of the vehicle. For instance, the attacker could spoof the victim AV's mmWave radar into \textit{seeing} a non-existent obstacle. Consequently, the victim AV changes its speed. 
\vspace{6pt}
\begin{figure}[htb!]
    \centering
    \includegraphics[trim={1.2cm 8cm 21cm 2.5cm},clip,width=0.85\columnwidth]{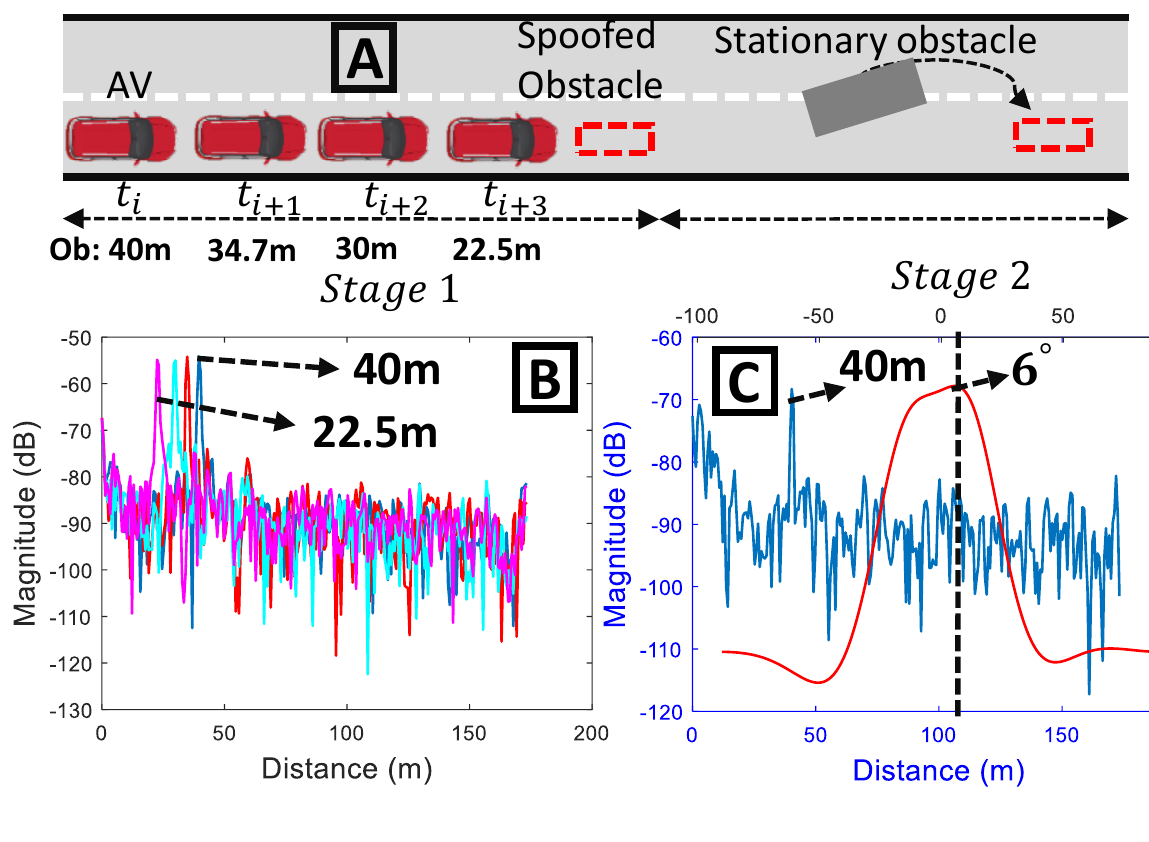}
    \vspace{-10pt}
    \caption{Illustration of scenario 5. Victim AV decelerates and accelerates to the set speed in response to the spoofed obstacle ahead before crashing on to the parked truck.}
    \vspace{3pt}
    \label{fig:scenario5}
\end{figure}
\begin{figure}[htb!]
    \centering
    \begin{minipage}[b]{0.49\columnwidth}
    \includegraphics[trim={5.5cm 9cm 5cm 10cm},clip,width=1\columnwidth]{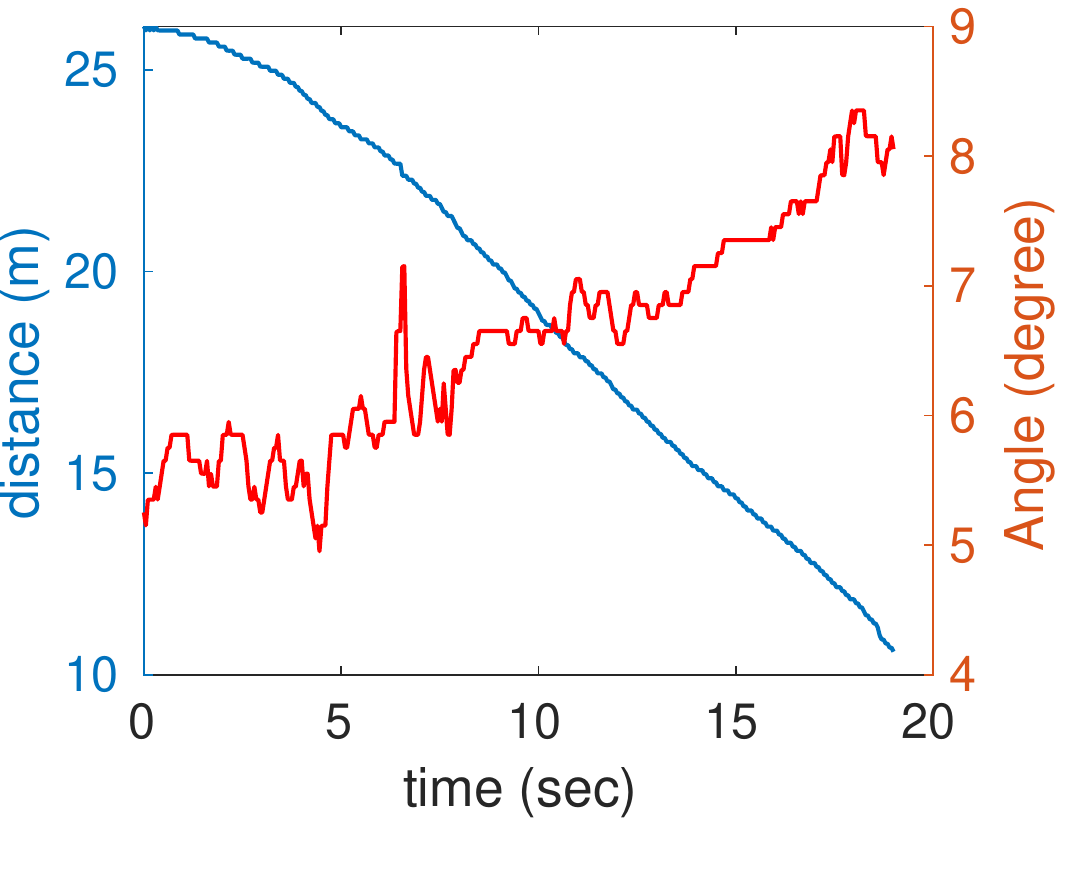}
    \caption{Distance/Angle of victim AV tracked by the attacker.}
    \label{fig:scenario5dist}
    \end{minipage}
    ~~
    \begin{minipage}[b]{0.45\columnwidth}
    \includegraphics[trim={5.5cm 9.5cm 5cm 10cm},clip,width=1.1\columnwidth]{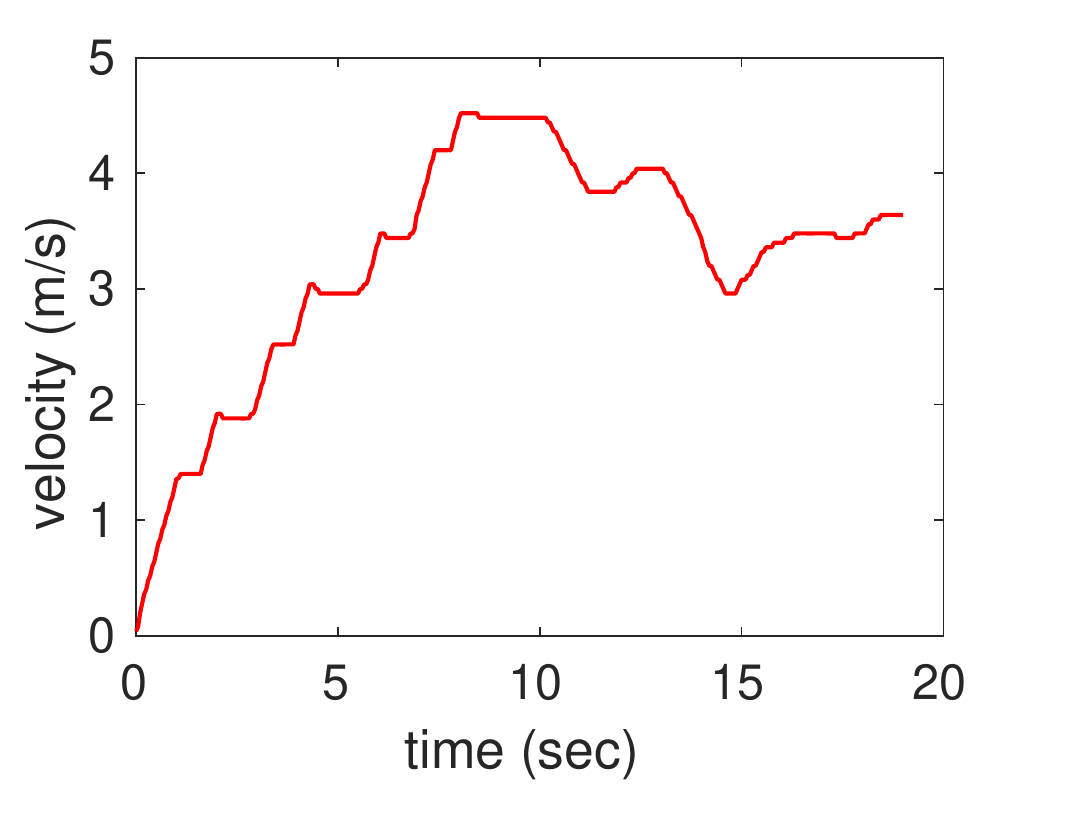}
    \caption{Velocity of victim AV tracked by the attacker.}
    \label{fig:scenario5vel}
    \end{minipage}
    \vspace{0pt}
\end{figure}

\begin{figure*}
    \centering
    \begin{minipage}[t]{0.47\textwidth}
    \centering
    \includegraphics[trim={4cm 10.5cm 19cm 3cm},clip,width=0.7\columnwidth]{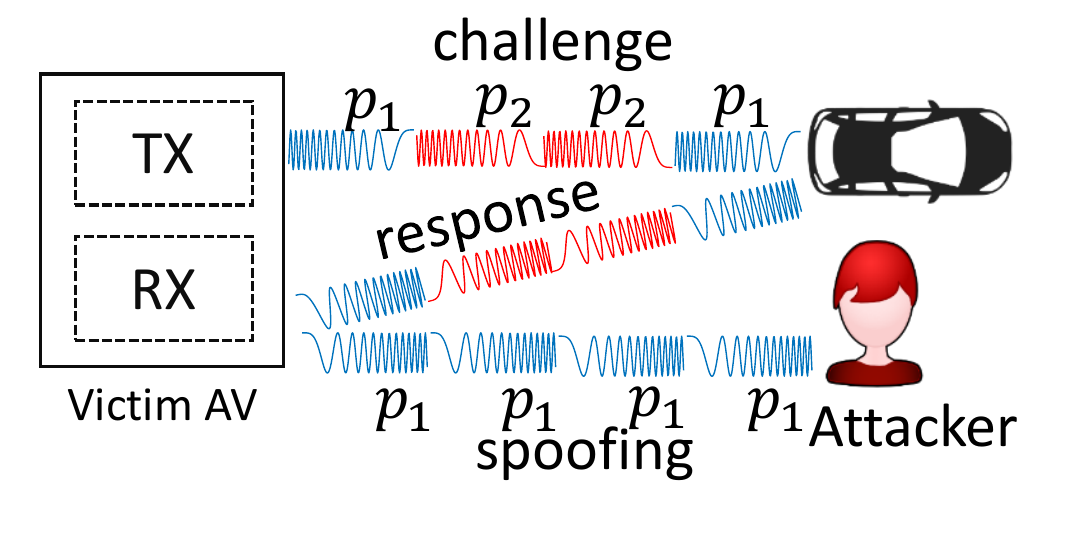}
    \vspace{-6pt}
    \caption{Illustration of \textit{challenge-response} spoofing detection where the victim AV transmits a challenge with parameters $p_1$ and $p_2$ and the reflected signal is used as a response.}
    \label{fig:challengeresponse}
    \end{minipage}
    ~
    \begin{minipage}[t]{0.25\textwidth}
    \includegraphics[trim={2cm 9cm 22cm 2cm},clip,width=0.9\columnwidth]{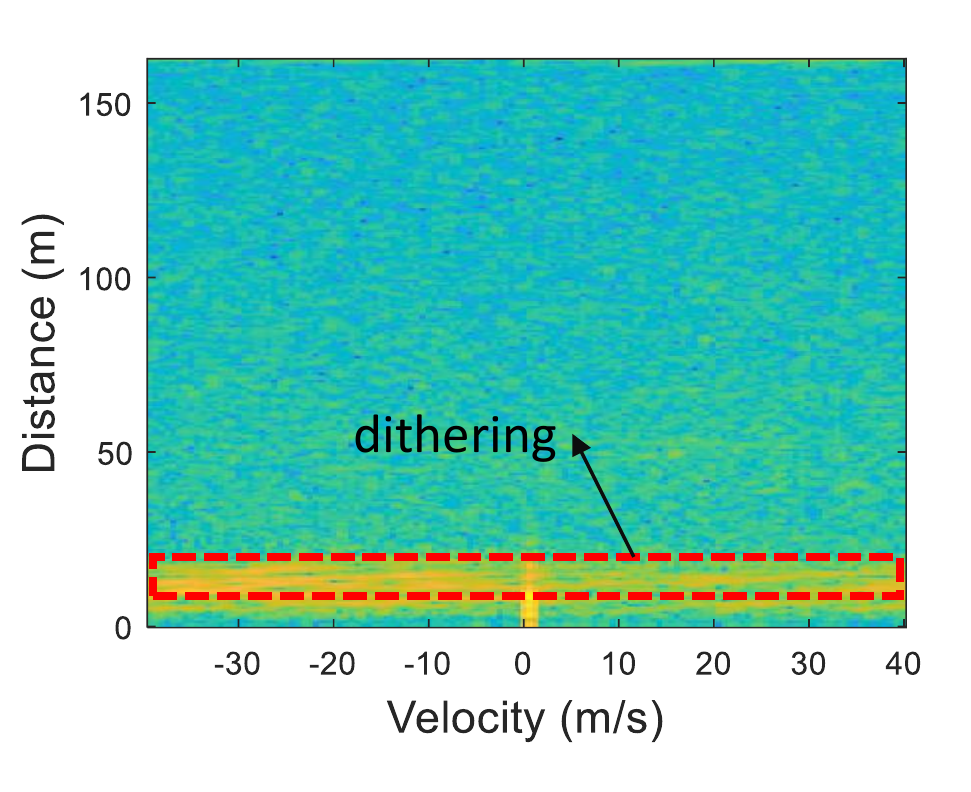}
    \caption{Spoofed obstacle is \textit{smeared} without distinct peak.}
    \label{fig:phasechallengeresponse}
    \end{minipage}
    ~
    \begin{minipage}[t]{0.23\textwidth}
    \includegraphics[trim={5.3cm 9cm 6cm 10cm},clip,width=0.9\columnwidth]{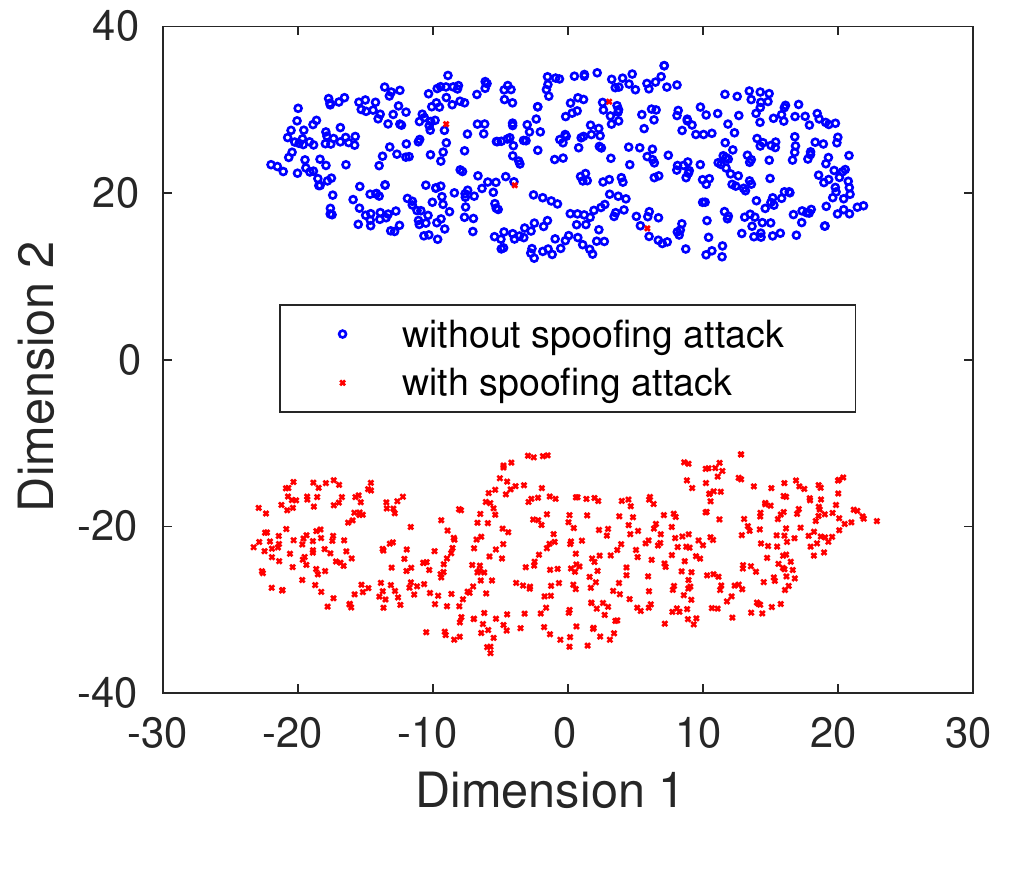}
    \caption{2-Dimensional representation of the features.}
    \label{fig:tsne}
    \end{minipage}
    \vspace{6pt}
\end{figure*}

\par 
Inspired by the above discussion on TACC safety, we construct the following scenario. Instead of a real physical vehicle ahead of the victim AV, the attackers spoof a fake leading vehicle. The objective of the attack is to influence the victim AV to decelerate and accelerate in response to a spoofing attack, leading to a crash, as shown in Block [A] in Fig. \ref{fig:scenario5}. At the beginning, the attackers track the position of the victim AV (Sec. \ref{sec:trackingsys}). Fig. \ref{fig:scenario5dist} shows the tracked position of the victim AV. Before the attack, the victim AV \textit{sees} no obstacle ahead of it in the decision making zone and maintains the set speed. At \textit{stage 1}, the attacker spoofs the victim AV with an obstacle at $40m$. In order to imitate a slow moving vehicle, the attacker continuously spoofs the victim AV with an obstacle at a decreasing distance. Block [B] in Fig. \ref{fig:scenario5} shows the spoofed obstacles at various distances from $40m$ to $22.5m$, which results in the victim AV decelerating as shown in Fig. \ref{fig:scenario5vel}. 
At \textit{stage 2}, the attacker stops the spoofing attack, which makes the AV start accelerating as it \textit{sees} no obstacle in front of it. In addition, the attackers employ the attack strategy provided in Sec. \ref{sec: asyncsngleattack} to fake the location of the stationary obstacle. The goal is to let the stationary obstacle appear out of the way of the victim AV. The attackers spoof the fake location of the stationary obstacle at 40m with respect to the victim AV and at an angle of $6^{\circ}$, as shown in block [C] in Fig. \ref{fig:scenario5}, which results in the stationary obstacle deviating by $~4m$ from the victim AV's lane. Fig. \ref{fig:scenario5vel} shows the speed timeline of the victim AV in the experiment. The victim AV accelerates and reaches a steady velocity before it decelerates due to the spoofed obstacle. At around $15$ sec, the victim AV accelerates again at which point the attackers spoof the stationary obstacle out of its way. The victim AV would have crashed on to the stationary obstacle at this point in time, leading to a fatal situation. 

\vspace{-6pt}
\section{Defending Strategies}
To defend the above fatal attacks, we propose two defending strategies to reliably detect spoofing attacks on AVs. 
\vspace{3pt}

\subsubsection{{\textbf{Challenge-Response}}}
Unlike LiDAR pulses that can be randomized to mitigate LiDAR spoofing attack, mmWave radar uses a well defined waveform designed to meet specific sensing capabilities \cite{Ti2018programming}. Such predictability empowers attackers to perform the attacks developed in this paper. The attackers can eavesdrop on the victims waveform and deduce its parameters to perform the attacks. Alternatively the attacker could simply record the signal from the victim AV and replay it at a later time to launch the attacks. To address this problem, departing from the traditional fixed waveform design, we propose to use a radar frame consisting of chirps with varying parameters, yet meet the design criteria. The transmitted chirp in Eq. \ref{eq:xt} is:
\begin{equation}\label{eq:txchirpwithphase}
    x(t) = cos\left(2\pi f_{start}t+\pi \frac{B}{T_{chirp}}t^2 + \phi_{init}(n)\right),
\end{equation}
where $n$ is the index of the $n^{th}$ chirp. The parameters $f_{start}$, $\frac{B}{T_{chirp}}$, and $\phi_{init}$ can be randomized across chirps or frames, thus mitigating the spoofing attacks. To that end, we introduce the \textit{challenge-response} authentication for mmWave radar, where the AV radar transmits chirp with randomized parameter as a challenge. The response, i.e., the received signal at the radar, is verified for authenticity. Fig. \ref{fig:challengeresponse} shows the AV transmitting a \textit{challenge} with chirps having random parameters $p_1$ and $p_2$. The parameter $p$ could be phase $\phi_{init}$ or $f_{start}$. 

In the experiment, we used $\phi_{init}$ and $f_{start}$ as two parameters. For the method with $\phi_{init}$ as a random parameter, the AV radar transmits a sequence of chirps in Eq. \ref{eq:txchirpwithphase} with a random phase $\phi_{init}$ for each transmitted chirp $n$. The attacker transmits a spoofing waveform with $\phi_{init} = 0$. Fig. \ref{fig:phasechallengeresponse} shows the radar detection of the victim AV. The attacker's spoofing signals fail to generate a peak. Instead, it appears as a smearing. In the second \textit{challenge-response} method, the AV varies $f_{start}$ randomly. Therefore, the attacker fails to register any meaningful spoofing effect on the victim AV. 

\vspace{3pt}
\subsubsection{{\textbf{Waveform fingerprinting}}}
Recently RF fingerprinting has gained attention in its ability to detect spoofing attacks. 
However, there are two challenges to utilize the RF fingerprinting techniques in radar. 
First, the uniqueness and non-linearity associated with the components of the transmitter hardware result in the RF features that are unique to each device. However, the radar transmitter shown in Fig. \ref{fig:systemoverview} is significantly simpler than communication devices, which leads to smaller feature set. 
Second, 
the received signal in radar is a superposition of the reflections from all the targets and the attackers. Those signals undergo two-way path loss in addition to reflection loss due to the reflecting object. 

To address the challenges, we find that the statistical characteristics of the AVs mmWave signal is influenced by the attackers waveform, which give us the opportunity to detect the attacks. Hence, we propose to use the statistical features, including: \textit{standard deviation, kurtosis, skewness} of magnitude and phase of the received signal \cite{joo2020hold}.
The amplitude of the received signal at radar is influenced by the two-way propagation path of the transmitted signal. One of the key requirement of RF fingerprinting is that the wireless channel influences need to be eliminated. Therefore, the received signal in Eq. \ref{eq:fb} is rms-normalized before feature extraction. The rms-normalization of Eq. \ref{eq:fb} is $    r_{rms}[n] = \frac{r[n]}{\sqrt{\frac{\sum_{n=1}^{N}r[n]^2}{N}}}
$,
where $N$ is the total number of samples in a received chirp.

\begin{figure}
        \centering
        \vspace{-3pt}
        \begin{minipage}[b]{0.35\columnwidth}
        \includegraphics[trim={6.5cm 10.3cm 7.5cm 11cm},clip,width=1\columnwidth]{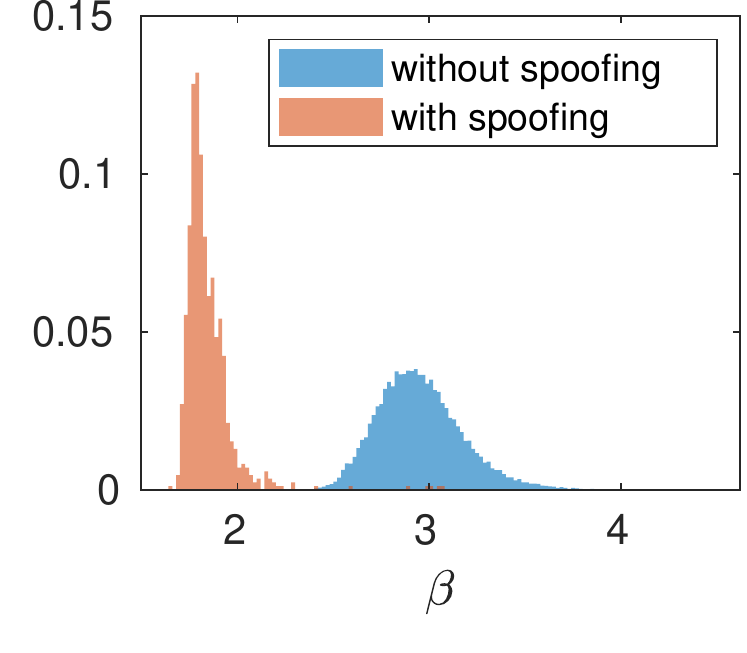}
        \end{minipage}
        \begin{minipage}[b]{0.35\columnwidth}
        \includegraphics[trim={6.5cm 10.3cm 7.5cm 11cm},clip,width=1\columnwidth]{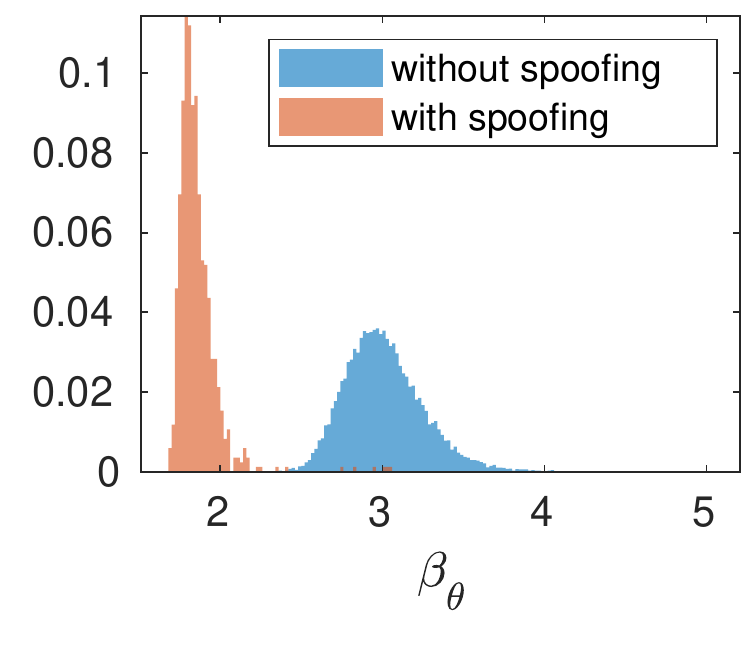}
        \end{minipage}\\
        \begin{minipage}[b]{0.35\columnwidth}
        \includegraphics[trim={6.5cm 10.3cm 7.5cm 11cm},clip,width=1\columnwidth]{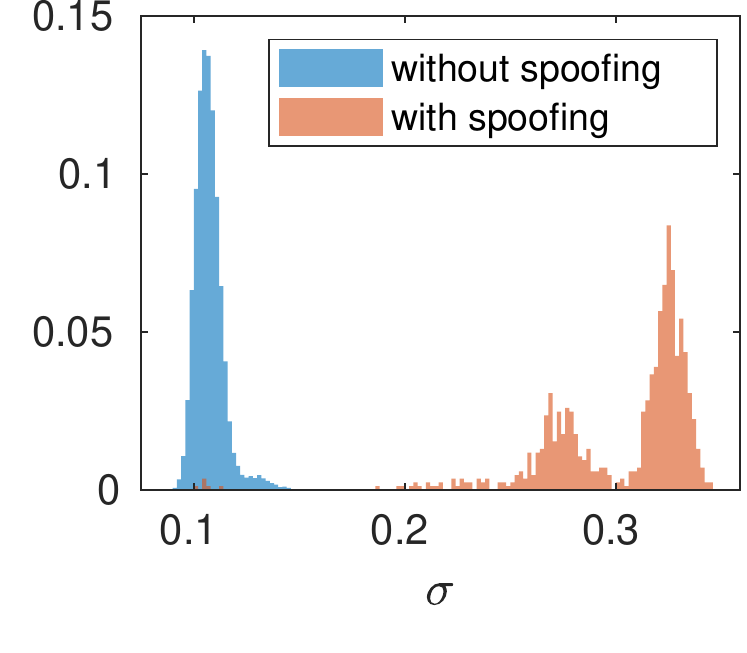}
        \end{minipage}
        \begin{minipage}[b]{0.35\columnwidth}
        \includegraphics[trim={6.5cm 10.3cm 7.5cm 11cm},clip,width=1\columnwidth]{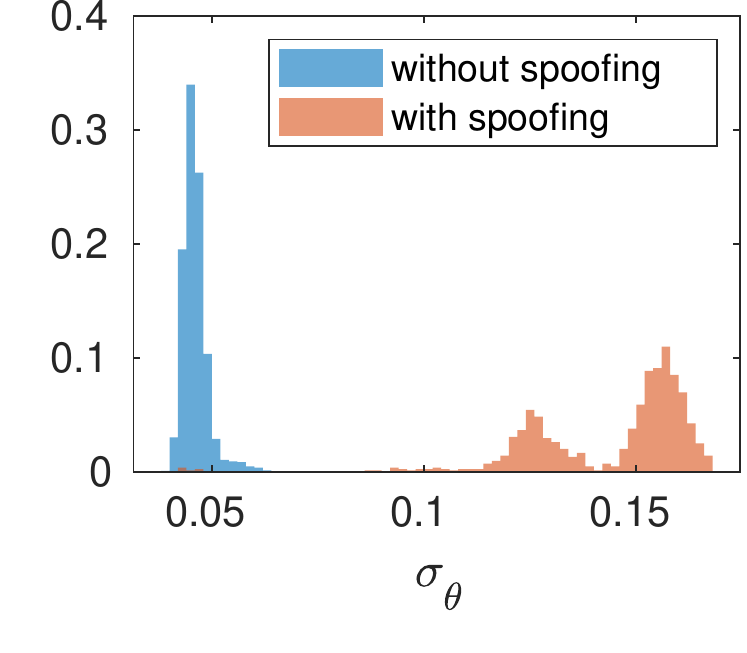}
        \end{minipage}
        \vspace{-9pt}
        \caption{Histogram of the received signal w/o spoofing.}
        \vspace{6pt}
        \label{fig:histogram}
    \end{figure}

The rms-normalized signal is then used to extract RF features. Since the AV mmWave signal does not use any specific preamble, we use the entire chirp of 256 samples to extract the features. Fig. \ref{fig:histogram} shows the distribution of the features when there is no spoofing attack and when the AVs mmWave radar is under spoofing attack. We can clearly see that the spoofing attacks change the distributions of the proposed features. As a result, the attacks can be reliably detected.

The AV software needs to detect spoofing attacks from the received signals. Only the training signals from the AV's own radar can be used. Therefore, we propose to use \textit{one-class SVM} as the machine learning framework to identify spoofing attack \cite{chalapathy2018anomaly}. We use 3000 received signals without spoofing attacks as training data. The testing data comprises 3000 signals under spoofing attack. Fig. \ref{fig:tsne} shows the 2-D embedding of the features. We can see distinct grouping between the features with (red points) and without (blue points) spoofing attack. The spoofing detection accuracy is $98.9\%$. 

\vspace{-6pt}
\section{Conclusion}
\vspace{-3pt}
In this paper, we perform the end-to-end security analysis of mmWave radars in AVs by designing and implementing practical physical layer attack and defense strategies in real-world scenarios. 
Two core attacking strategies are rigorously designed and validated, which can reliably add spoofing obstacles or fake the locations of existing obstacles. Based on the state-of-the-art mmWave and AV testbed, the attackers combine the core attacking strategies to continuously spoof the AV into making hazardous safety-critical driving decisions, leading to fatal accidents. 
To defend such attacks, we propose two defense strategies and prove the defending solutions can detect the sensor spoofing attacks with very high accuracy. 

\section*{Acknowledgment}
The authors would like to thank Roman Dmowski for assisting in the autonomous vehicle field experiments.


\vspace{-3pt}
\appendix
\subsection{Modules of Baidu Apollo System}
\label{sec:softwaremodules}
\vspace{-3pt}
\paragraph*{\textbf{ROI filter}} Region of interest (ROI) is the region that lies in the AV's driving path, which directly influences the AV's driving decisions. 
The detected objects should lie within the ROI, while objects outside ROI are discarded. 

\paragraph*{\textbf{Object Tracking}}
\label{sec:KF}
Once the obstacles are filtered based on ROI, they are tracked using the Kalman filter.

\paragraph*{\textbf{Track matching}} Objects sensed by the sensors in the current sensing cycle are associated to the objects sensed in the previous cycle using data matching algorithms. 
Apollo Baidu matches the objects detected in a current radar frame to the existing tracks using gated Hungarian matching algorithm. When the object in the current radar frame does not match any existing tracks, new track is created for that object. 

\paragraph*{\textbf{Path planning}} \label{sec:pathplanning}
AVs typically use a search algorithm with a certain cost function to decide the best path that an AV follows \cite{fan2018baidu}. 
In Apollo Baidu, the cost function is $C_{total} = C_{env} + C_{obstacle}$ \cite{fan2018baidu}. The obstacle cost $C_{obstacle}$ is given by
\vspace{-6pt}
\begin{equation}\label{eq:costfunc}
    C_{obstacle} = \begin{cases} 
    0 & d > d_{n}, \\
    C_{nudge}(d-d_{c}) & d_c \leq d \leq d_n, \\
    C_{collision} & d < d_c.
    \end{cases}
\end{equation}

\paragraph*{\textbf{Baidu Apollo Parameters}} \label{sec:BaiduApolloParam}
Important parameters used in Baidu Apollo AV software are summarized in Table \ref{table:Baiduparam}.
\begin{table}[htb!]
\vspace{6pt}
\centering
\begin{tabular}{|c | c | p{35mm} |} 
 \hline
 Parameter & Value & Description\\ [0.5ex] 
 \hline
 min-lane-change-length & 5m & minimum distance to change a lane\\ 
 min-lane-change-prepare-length & 60m & minimum distance to prepare a lane change \\
 follow-min-distance & 3m & minimum distance to follow an obstacle\\
 min-stop-distance-obstacle & 6m & minimum stop distance from in-lane obstacle\\
 max-stop-distance-obstacle & 10m & maximum stop distance from in-lane obstacles \\
 lane-change-prepare-length & 80m & distance to prepare for lane change \\
 min-lane-change-prepare-length & 10m & minimum distance to prepare for lane change \\
 min-nudge-distance & 0.2m & minimum distance to nudge \\
 max-nudge-distance & 1.1m & maximum distance to nudge \\
  min-yield-distance & 5m & minimum distance to yield\\ [1ex] 
 \hline
\end{tabular}
\vspace{6pt}
\caption{Parameters used by the Apollo Baidu's perception, planning, and navigation modules.}
\vspace{6pt}
\label{table:Baiduparam}
\end{table}

\bibliographystyle{IEEEtran}
\bibliography{sample-base}

\begin{thebibliography}{10}
\providecommand{\url}[1]{#1}
\csname url@samestyle\endcsname
\providecommand{\newblock}{\relax}
\providecommand{\bibinfo}[2]{#2}
\providecommand{\BIBentrySTDinterwordspacing}{\spaceskip=0pt\relax}
\providecommand{\BIBentryALTinterwordstretchfactor}{4}
\providecommand{\BIBentryALTinterwordspacing}{\spaceskip=\fontdimen2\font plus
\BIBentryALTinterwordstretchfactor\fontdimen3\font minus
  \fontdimen4\font\relax}
\providecommand{\BIBforeignlanguage}[2]{{%
\expandafter\ifx\csname l@#1\endcsname\relax
\typeout{** WARNING: IEEEtran.bst: No hyphenation pattern has been}%
\typeout{** loaded for the language `#1'. Using the pattern for}%
\typeout{** the default language instead.}%
\else
\language=\csname l@#1\endcsname
\fi
#2}}
\providecommand{\BIBdecl}{\relax}
\BIBdecl

\bibitem{litman2017autonomous}
T.~Litman, \emph{Autonomous vehicle implementation predictions}.\hskip 1em plus
  0.5em minus 0.4em\relax Victoria Transport Policy Institute Victoria, Canada,
  2017.

\bibitem{ApolloBaidu}
\url{https://github.com/ApolloAuto/apollo}.

\bibitem{kato2018autoware}
S.~Kato, S.~Tokunaga, Y.~Maruyama, S.~Maeda, M.~Hirabayashi, Y.~Kitsukawa,
  A.~Monrroy, T.~Ando, Y.~Fujii, and T.~Azumi, ``Autoware on board: Enabling
  autonomous vehicles with embedded systems,'' in \emph{2018 ACM/IEEE 9th
  International Conference on Cyber-Physical Systems (ICCPS)}.\hskip 1em plus
  0.5em minus 0.4em\relax IEEE, 2018, pp. 287--296.

\bibitem{openpilot}
``"openpilot: Open source driving agent",''
  \url{https://github.com/commaai/openpilot}.

\bibitem{teslaAccidentMay2016}
A.~Singhvi and K.~Russell, ``Inside the self-driving tesla fatal accident,''
  \url{https://www.nytimes.com/interactive/2016/07/01/business/insidetesla-accident.html}.

\bibitem{Tesla2018Accident}
\url{https://www.bbc.com/news/world-us-canada-43604440}.

\bibitem{Teslaaccident}
``Rear-end collision between a car operating with advanced driver assistance
  systems and a stationary fire truck, culver city, california, january 22,
  2018,'' \url{https://dms.ntsb.gov/public/62500-62999/62683/627686.pdf}.

\bibitem{petit2015remote}
J.~Petit, B.~Stottelaar, M.~Feiri, and F.~Kargl, ``Remote attacks on automated
  vehicles sensors: Experiments on camera and lidar,'' \emph{Black Hat Europe},
  vol.~11, p. 2015, 2015.

\bibitem{shin2017illusion}
H.~Shin, D.~Kim, Y.~Kwon, and Y.~Kim, ``Illusion and dazzle: Adversarial
  optical channel exploits against lidars for automotive applications,'' in
  \emph{International Conference on Cryptographic Hardware and Embedded
  Systems}.\hskip 1em plus 0.5em minus 0.4em\relax Springer, 2017, pp.
  445--467.

\bibitem{cao2019adversarial}
Y.~Cao, C.~Xiao, B.~Cyr, Y.~Zhou, W.~Park, S.~Rampazzi, Q.~A. Chen, K.~Fu, and
  Z.~M. Mao, ``Adversarial sensor attack on lidar-based perception in
  autonomous driving,'' in \emph{Proceedings of the 2019 ACM SIGSAC Conference
  on Computer and Communications Security}, 2019, pp. 2267--2281.

\bibitem{chauhan2014platform}
R.~Chauhan, ``A platform for false data injection in frequency modulated
  continuous wave radar,'' 2014.

\bibitem{yan2016can}
C.~Yan, W.~Xu, and J.~Liu, ``Can you trust autonomous vehicles: Contactless
  attacks against sensors of self-driving vehicle,'' \emph{DEF CON}, vol.~24,
  2016.

\bibitem{niSDR}
``Introduction to the ni mmwave transceiver system hardware,''
  \url{https://www.ni.com/en-us/innovations/white-papers/16/introduction-to-the-ni-mmwave-transceiver-system-hardware.html}.

\bibitem{cao2019adversarialOld}
Y.~Cao, C.~Xiao, D.~Yang, J.~Fang, R.~Yang, M.~Liu, and B.~Li, ``Adversarial
  objects against lidar-based autonomous driving systems,'' \emph{arXiv
  preprint arXiv:1907.05418}, 2019.

\bibitem{ranganathan2012physical}
A.~Ranganathan, B.~Danev, A.~Francillon, and S.~Capkun, ``Physical-layer
  attacks on chirp-based ranging systems,'' in \emph{Proceedings of the fifth
  ACM conference on Security and Privacy in Wireless and Mobile Networks},
  2012, pp. 15--26.

\bibitem{chauhan2014demonstration}
R.~Chauhan, R.~M. Gerdes, and K.~Heaslip, ``Demonstration of a false-data
  injection attack against an fmcw radar,'' \emph{12th escar Europe}, p. 135,
  2014.

\bibitem{miura2019low}
N.~Miura, T.~Machida, K.~Matsuda, M.~Nagata, S.~Nashimoto, and D.~Suzuki, ``A
  low-cost replica-based distance-spoofing attack on mmwave fmcw radar,'' in
  \emph{Proceedings of the 3rd ACM Workshop on Attacks and Solutions in
  Hardware Security Workshop}, 2019, pp. 95--100.

\bibitem{jha2019kayotee}
S.~Jha, T.~Tsai, S.~Hari, M.~Sullivan, Z.~Kalbarczyk, S.~W. Keckler, and R.~K.
  Iyer, ``Kayotee: A fault injection-based system to assess the safety and
  reliability of autonomous vehicles to faults and errors,'' \emph{arXiv
  preprint arXiv:1907.01024}, 2019.

\bibitem{jha2019ml}
S.~Jha, S.~Banerjee, T.~Tsai, S.~K. Hari, M.~B. Sullivan, Z.~T. Kalbarczyk,
  S.~W. Keckler, and R.~K. Iyer, ``Ml-based fault injection for autonomous
  vehicles: a case for bayesian fault injection,'' in \emph{2019 49th Annual
  IEEE/IFIP International Conference on Dependable Systems and Networks
  (DSN)}.\hskip 1em plus 0.5em minus 0.4em\relax IEEE, 2019, pp. 112--124.

\bibitem{rubaiyat2018experimental}
A.~H.~M. Rubaiyat, Y.~Qin, and H.~Alemzadeh, ``Experimental resilience
  assessment of an open-source driving agent,'' in \emph{2018 IEEE 23rd Pacific
  Rim International Symposium on Dependable Computing (PRDC)}.\hskip 1em plus
  0.5em minus 0.4em\relax IEEE, 2018, pp. 54--63.

\bibitem{sitawarin2018darts}
C.~Sitawarin, A.~N. Bhagoji, A.~Mosenia, M.~Chiang, and P.~Mittal, ``Darts:
  Deceiving autonomous cars with toxic signs,'' \emph{arXiv preprint
  arXiv:1802.06430}, 2018.

\bibitem{chernikova2019self}
A.~Chernikova, A.~Oprea, C.~Nita-Rotaru, and B.~Kim, ``Are self-driving cars
  secure? evasion attacks against deep neural networks for steering angle
  prediction,'' in \emph{2019 IEEE Security and Privacy Workshops (SPW)}.\hskip
  1em plus 0.5em minus 0.4em\relax IEEE, 2019, pp. 132--137.

\bibitem{jia2019fooling}
Y.~Jia, Y.~Lu, J.~Shen, Q.~A. Chen, H.~Chen, Z.~Zhong, and T.~Wei, ``Fooling
  detection alone is not enough: Adversarial attack against multiple object
  tracking,'' in \emph{International Conference on Learning Representations},
  2019.

\bibitem{boloor2020attacking}
A.~Boloor, K.~Garimella, X.~He, C.~Gill, Y.~Vorobeychik, and X.~Zhang,
  ``Attacking vision-based perception in end-to-end autonomous driving
  models,'' \emph{Journal of Systems Architecture}, p. 101766, 2020.

\bibitem{shoukry2015pycra}
Y.~Shoukry, P.~Martin, Y.~Yona, S.~Diggavi, and M.~Srivastava, ``Pycra:
  Physical challenge-response authentication for active sensors under spoofing
  attacks,'' in \emph{Proceedings of the 22nd ACM SIGSAC Conference on Computer
  and Communications Security}, 2015, pp. 1004--1015.

\bibitem{schneider2005automotive}
M.~Schneider, ``Automotive radar-status and trends,'' in \emph{German microwave
  conference}, 2005, pp. 144--147.

\bibitem{vubiq42dbi}
``Wr-15 waveguide horn antenna operating from 50 ghz to 75 ghz with a nominal
  42 dbi gain with ug-385/u round cover flange,''
  \url{https://www.pasternack.com/images/ProductPDF/PE9881-42.pdf}.

\bibitem{abdelaziz2016security}
A.~Abdelaziz, C.~E. Koksal, and H.~El~Gamal, ``On the security of angle of
  arrival estimation,'' in \emph{2016 IEEE Conference on Communications and
  Network Security (CNS)}.\hskip 1em plus 0.5em minus 0.4em\relax IEEE, 2016,
  pp. 109--117.

\bibitem{LincolnMKZ}
\url{https://ubwp.buffalo.edu/cavas/introduction-lincoln/}.

\bibitem{dmowski2019research}
R.~Dmowski, S.~Korzelius, J.~Huang, X.~Liu, F.~Hajiaghajani,
  G.~Balasubramaniam, S.~Ravi, R.~V. Soni, A.~Gupta, A.~Gupta \emph{et~al.},
  ``Research at the intersection of big data and connected and autonomous
  vehicles,'' 2019.

\bibitem{TI6843}
``Ti iwr6843 single-chip 60- to 64-ghz mmwave sensor (rev. b),''
  \url{https://www.ti.com/document-viewer/IWR6843/datasheet/description-x5103#x5103}.

\bibitem{Ti2018programming}
V.~Dham, ``Programming chirp parameters in ti radar devices,''
  \emph{Application Report SWRA553, Texas Instruments}, 2020.

\bibitem{joo2020hold}
K.~Joo, W.~Choi, and D.~H. Lee, ``Hold the door! fingerprinting your car key to
  prevent keyless entry car theft,'' \emph{arXiv preprint arXiv:2003.13251},
  2020.

\bibitem{chalapathy2018anomaly}
R.~Chalapathy, A.~K. Menon, and S.~Chawla, ``Anomaly detection using one-class
  neural networks,'' \emph{arXiv preprint arXiv:1802.06360}, 2018.

\bibitem{fan2018baidu}
H.~Fan, F.~Zhu, C.~Liu, L.~Zhang, L.~Zhuang, D.~Li, W.~Zhu, J.~Hu, H.~Li, and
  Q.~Kong, ``Baidu apollo em motion planner,'' \emph{arXiv preprint
  arXiv:1807.08048}, 2018.

\end{thebibliography}
%

\end{document}